\documentclass[pdflatex,sn-mathphys-num]{sn-jnl}

\usepackage{caption}
\usepackage{subcaption}
\usepackage{pdflscape}

\usepackage{graphicx}%
\usepackage{multirow}%
\usepackage{amsmath,amssymb,amsfonts}%
\usepackage{amsthm}%
\usepackage{mathrsfs}%
\usepackage[title]{appendix}%
\usepackage{xcolor}%
\usepackage{textcomp}%
\usepackage{manyfoot}%
\usepackage{booktabs}%
\usepackage{algorithm}%
\usepackage{algorithmicx}%
\usepackage{algpseudocode}%
\usepackage{listings}%

\begin{document}

\title[Meteorite Record Survival Bias]{Perihelion history and atmospheric survival as primary drivers of the Earth’s meteorite record}

\author*[1]{\fnm{Patrick M.} \sur{Shober}}\email{planetarypat@gmail.com}

\author[2,3]{\fnm{Hadrien A.R.} \sur{Devillepoix}}

\author[1]{\fnm{Jeremie} \sur{Vaubaillon}}

\author[1,4]{\fnm{Simon} \sur{Anghel}}

\author[2,3]{\fnm{Sophie E.} \sur{Deam}}

\author[3,2]{\fnm{Eleanor K.} \sur{Sansom}}

\author[1]{\fnm{Francois} \sur{Colas}}

\author[5,1]{\fnm{Brigitte} \sur{Zanda}}

\author[6]{\fnm{Pierre} \sur{Vernazza}}

\author[2]{\fnm{Phil} \sur{Bland}}

\affil*[1]{\orgdiv{LTE}, \orgname{Observatoire de Paris, Université PSL, Sorbonne Université, Université de Lille, LNE, CNRS}, \orgaddress{\street{61 Avenue de l’Observatoire}, \city{Paris}, \postcode{75014}, \country{France}}}

\affil[2]{\orgdiv{Space Science and Technology Centre}, \orgname{Curtin University}, \orgaddress{\street{Kent Street}, \city{Perth}, \postcode{6102}, \state{Western Australia}, \country{Australia}}}

\affil[3]{\orgdiv{International Centre for Radio Astronomy Research}, \orgname{Curtin University}, \orgaddress{\street{Kent Street}, \city{Perth}, \postcode{6102}, \state{Western Australia}, \country{Australia}}}

\affil[4]{\orgname{Astronomical Institute of the Romanian Academy}, \orgaddress{\street{Cutitul de Argint 5}, \city{Bucharest}, \postcode{040557}, \country{Romania}}}

\affil[5]{\orgdiv{Institut de Minéralogie, Physique des Matériaux et Cosmochimie}, \orgname{Muséum National d’Histoire Naturelle, CNRS}, \city{Paris}, \postcode{75005}, \country{France}}

\affil[6]{\orgdiv{Laboratoire d’Astrophysique de Marseille}, \orgname{Aix-Marseille University, CNRS, CNES, LAM, Institut Origines},\orgaddress{\street{38 rue Frederic Joliot Curie}, \city{Marseille}, \postcode{13388}, \country{France}}}




\maketitle


\textbf{Models predict that more than half of all impacting meteoroids should be carbonaceous, reflecting the abundance of carbon-rich asteroids in the main belt and near-Earth space. Yet carbonaceous chondrites represent only about 4\% of meteorites recovered worldwide. Here, we analyse 7,982 meteoroid impacts and 540 potential meteorite falls from 19 global observation networks and demonstrate that intense thermal stress at low perihelion distances coupled with the filtering effect of Earth’s atmosphere explains this mismatch. Meteoroids repeatedly subjected to intense thermal cycling near the Sun fracture and weaken, removing the most friable objects even before atmospheric entry. Our data also show that tidally disrupted meteoroid streams produce especially fragile fragments that rarely survive to the ground. Consequently, compact, higher-strength, thermally-cycled bodies dominate the meteorite record. These findings reconcile the predicted carbonaceous flux with its scarcity in collections, underscoring how orbital evolution and atmospheric filtering shape the materials that reach Earth’s surface.} \\

Shooting stars in the night sky have always captivated human curiosity, yet only within the last century have we developed the capability to continuously monitor and begin to understand the terrestrial impact population by observing the ablation of small bodies in our atmosphere \cite{ceplecha1998meteor}. This transition from mere observation to detailed characterization has been driven by advancements in the sensitivity and cost of camera technologies that have rendered global monitoring both feasible and economical \cite{devillepoix2020global,colas2020fripon,vida2021global,borovivcka2022_one}. Today, comprehensive datasets from expanding fireball observation networks across the globe offer unprecedented insights into the centimetre to several-metre debris environment near the Earth -- a fundamental component in unravelling the history and mechanics of our solar system.

Meteorites, the remnants of these celestial encounters, provide unparalleled information about the origins and evolution of our solar system. Each fragment is a direct sample from a distant planetary body, offering clues to the ancient materials that coalesced and condensed from the protoplanetary disk, forming our solar system. However, to fully leverage this dataset, it is crucial to understand the biases introduced by Earth’s atmosphere — what objects are missing from our datasets? When we look at the proportions of different meteorite types found in our global collections, we find that the relative distribution is not at all consistent with that of the variety of asteroid spectral types we observe telescopically in the main asteroid belt or near-Earth space \citep{vernazza2008compositional}. Only very recently has this disconnect between the ordinary chondrites and their corresponding asteroid types been resolved through the identification of young asteroid families responsible for the majority of the meteorites we have in our collections \citep{brovz2024young,marsset2024massalia}. C- and B-type asteroids, with carbonaceous-like compositions, dominate the main belt and make up about 64\% of the main-belt's mass. Meanwhile, S-type asteroids, analogous to ordinary chondrites, account for approximately 8\% \citep{demeo2013taxonomic}. Despite this, only $\sim$4\% of the world's meteorite collections, comprising over 83,000 samples, are carbonaceous. This is still extraordinarily low, considering that the most accurate recent models have estimated that over 50\% of the impact flux at the top of the atmosphere within the meteorite-dropping size range should be carbonaceous \citep{brovz2024source}. Often, the friability of the carbonaceous material is used as a generally acceptable explanation for its paucity in the meteorite collections -- it is simply too weak to make it through our thick atmosphere \citep{mcmullan2024winchcombe,brovz2024source}. However, no one has ever been able to confirm this hypothesis, as a significant number of impact observations are necessary to get statistically significant results. 

Historically, the rarity of observations of meteorite-causing fireballs and the immense spatial coverage required to track them left many questions about this filter unanswered. However, several global meteor and fireball observation networks today cover a few percent of the Earth's surface. To characterize the atmospheric filter of meteoritic debris, we needed to combine the observations of 19 different observation networks spread across 39 countries \cite{kornos2014edmond,rudawska2015independent,jenniskens2016established,devillepoix2020global,colas2020fripon,vida2021global,borovivcka2022_one}. This collection of decades of observations reveal the clearest picture yet of the orbital distribution of impacting materials, and what is failing to make it through Earth's first line of defence --- the atmosphere. 

\section{Results}

\subsection{The top of the atmosphere population}\label{sec1}

\begin{figure}
    \centering
    \includegraphics[width=1\linewidth]{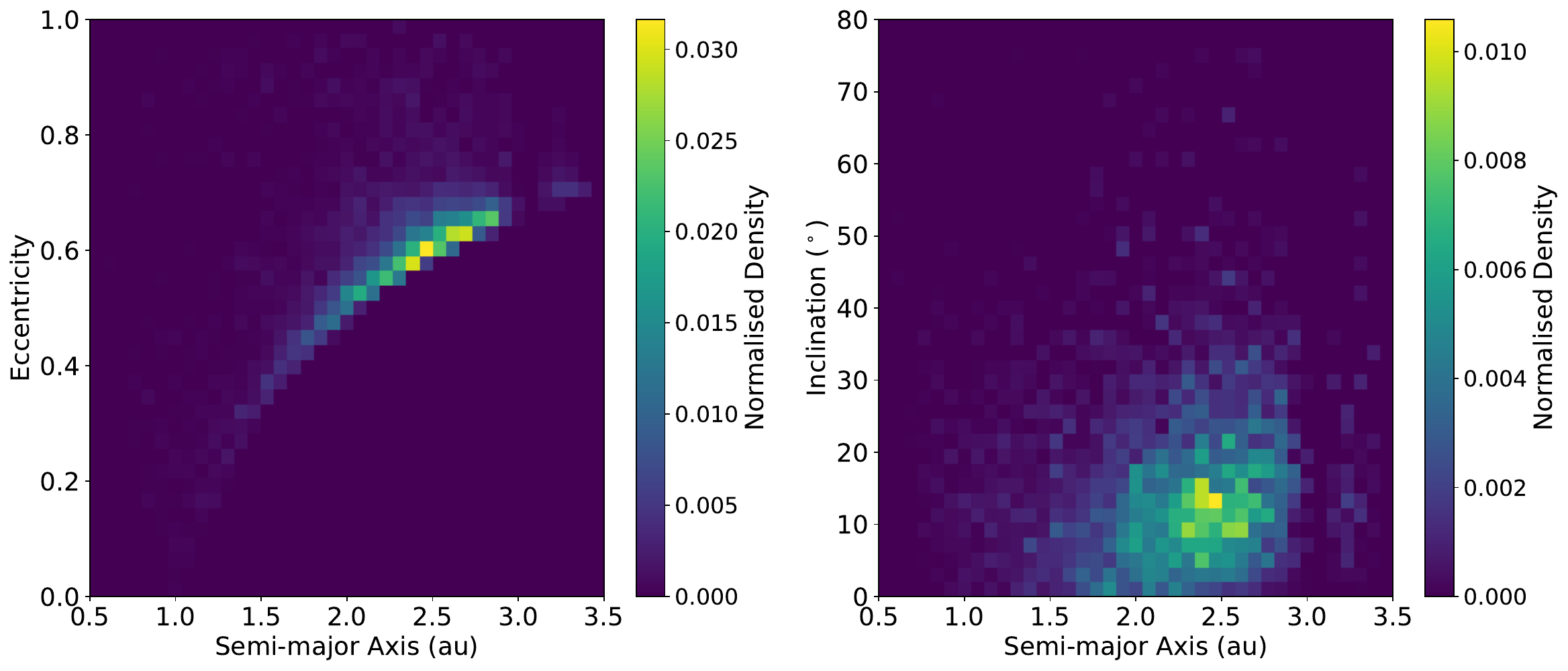}
    \caption{Orbital distribution of 7,982 impacts detected by EDMOND, CAMS, GMN, FRIPON, and EFN networks. Each impact is estimated to be equal to or greater than 10\,g (diameter $\gtrapprox$2\,cm) at the top of the atmosphere. Orbital distribution is normalized to the impact probability, as calculated using the method of \citet{pokorny2013opik}. Despite this, there still exists a concentration of meteoroids on orbits with q$\sim$1\,au.}
    \label{fig:aei_debiased}
\end{figure}
\phantomsection
\label{fig:aei_debiased}

The top-of-the-atmosphere population of meteoroids was identified through observations of 7,982 impacts detected by the EDMOND, CAMS, GMN, FRIPON, and EFN networks. The minimum approximate initial mass considered in the study was approximately 10\,g, and the most significant objects are metre-scale impactors. Critically, the observational bias was
removed from the top-of-the-atmosphere dataset by cutting the size-frequency distribution (SFD), where the slope started to significantly deviate from linear in log-log space (see Methods). Without removing this detection bias, the higher velocity component of the impact database will be significantly overestimated. Sensors are limited by the limiting brightness they can detect, and this magnitude is a function of the mass and the velocity \citep{ceplecha1998meteor}. Thus, the low-mass end range of the detectable population by a fireball observation network is over-represented with high velocities. In addition to this velocity bias, the impact probabilities with the Earth must be considered to accurately estimate the orbital distribution of centimetre-metre scale meteoroids in near-Earth space. As seen in Fig.~1, after removing these biases, we are left with a dataset of 7,982 impacts to estimate the orbital distribution of steady-state meteoroid flux in the centimetre to metre range. This distribution is still concentrated towards low-perihelion and low-inclination, indicating a genuine concentration towards these orbits unrelated to network biases or impact probabilities. 

\begin{figure}
    \centering
    \includegraphics[width=0.5\linewidth]{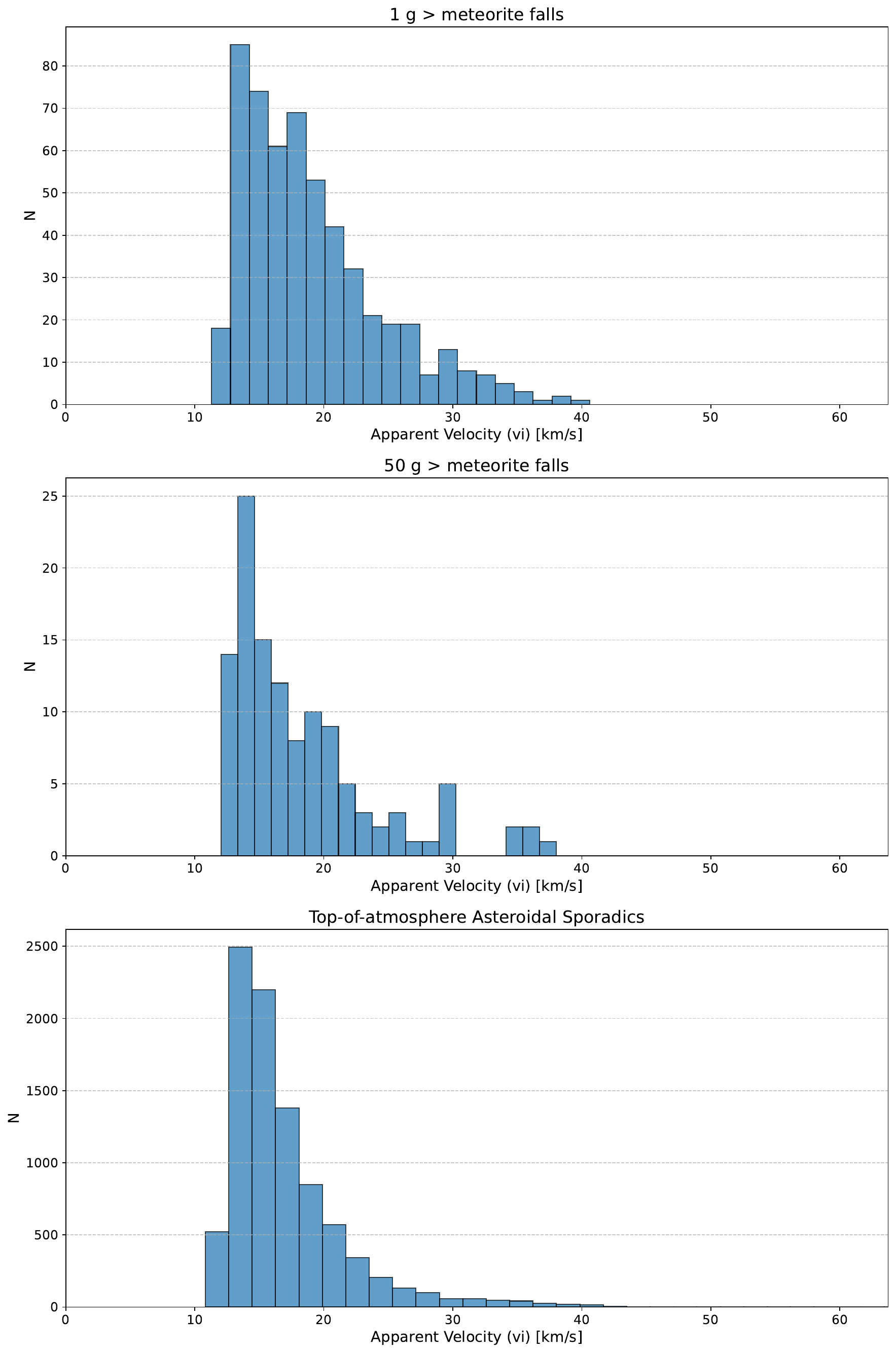}
    \caption{Apparent initial velocity distribution for the sporadic asteroidal top-of-the-atmosphere population versus the meteorite fall populations. The top-of-the-atmosphere population is derived from a debiased subset of observations from the EDMOND, CAMS, GMN, FRIPON, and EFN impact databases. The meteorite fall observations are taken from the GFO, EFN, and FRIPON datasets.}
    \label{fig:initial_velocities}
\end{figure}

\begin{figure}
    \centering
    \begin{subfigure}[b]{0.5\textwidth}
        \centering
        \includegraphics[width=\textwidth]{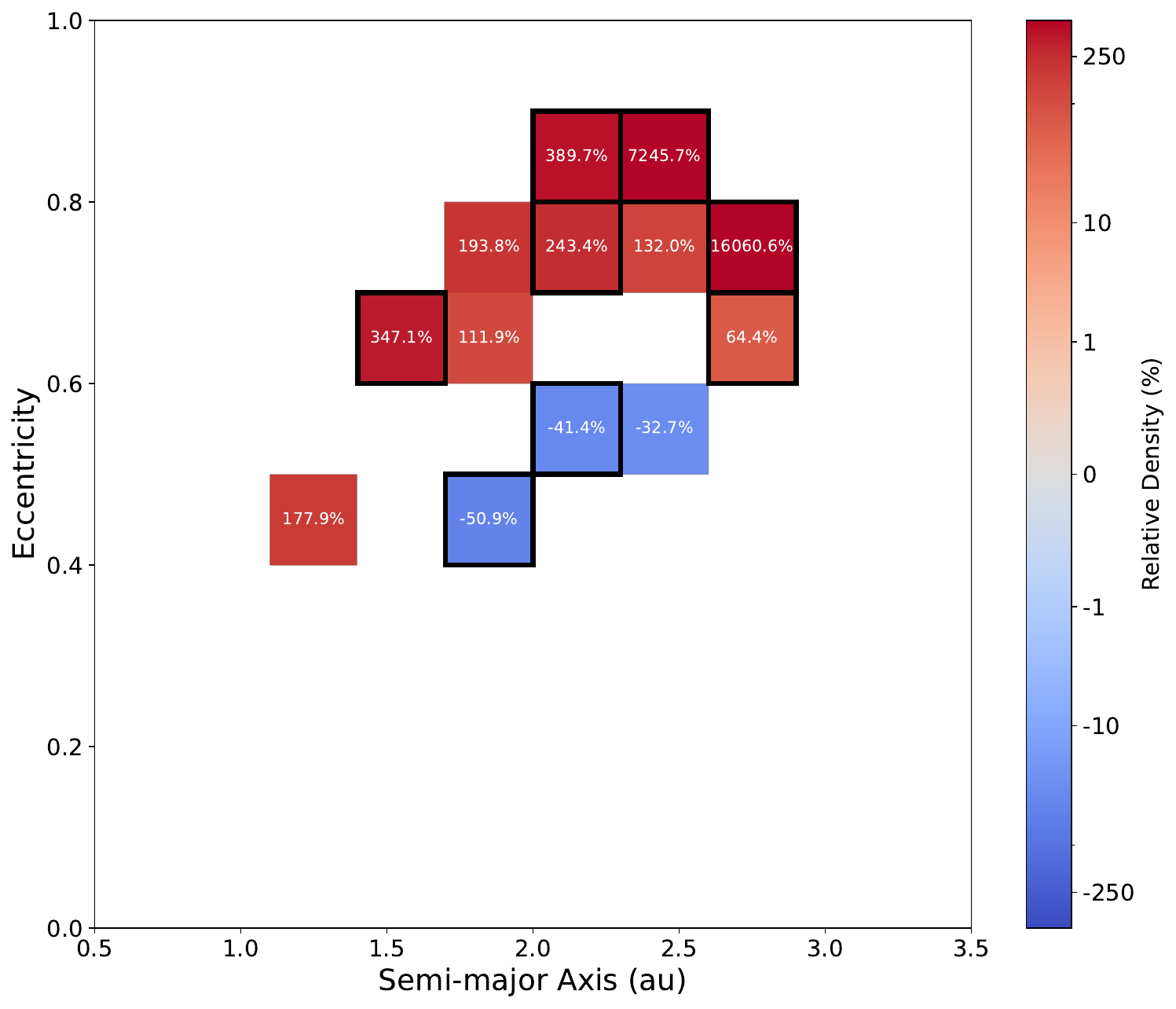} 
        \label{fig:image1}
    \end{subfigure}
    \hspace{-0.1\textwidth} 
    \begin{subfigure}[b]{0.5\textwidth}
        \centering
        \includegraphics[width=\textwidth]{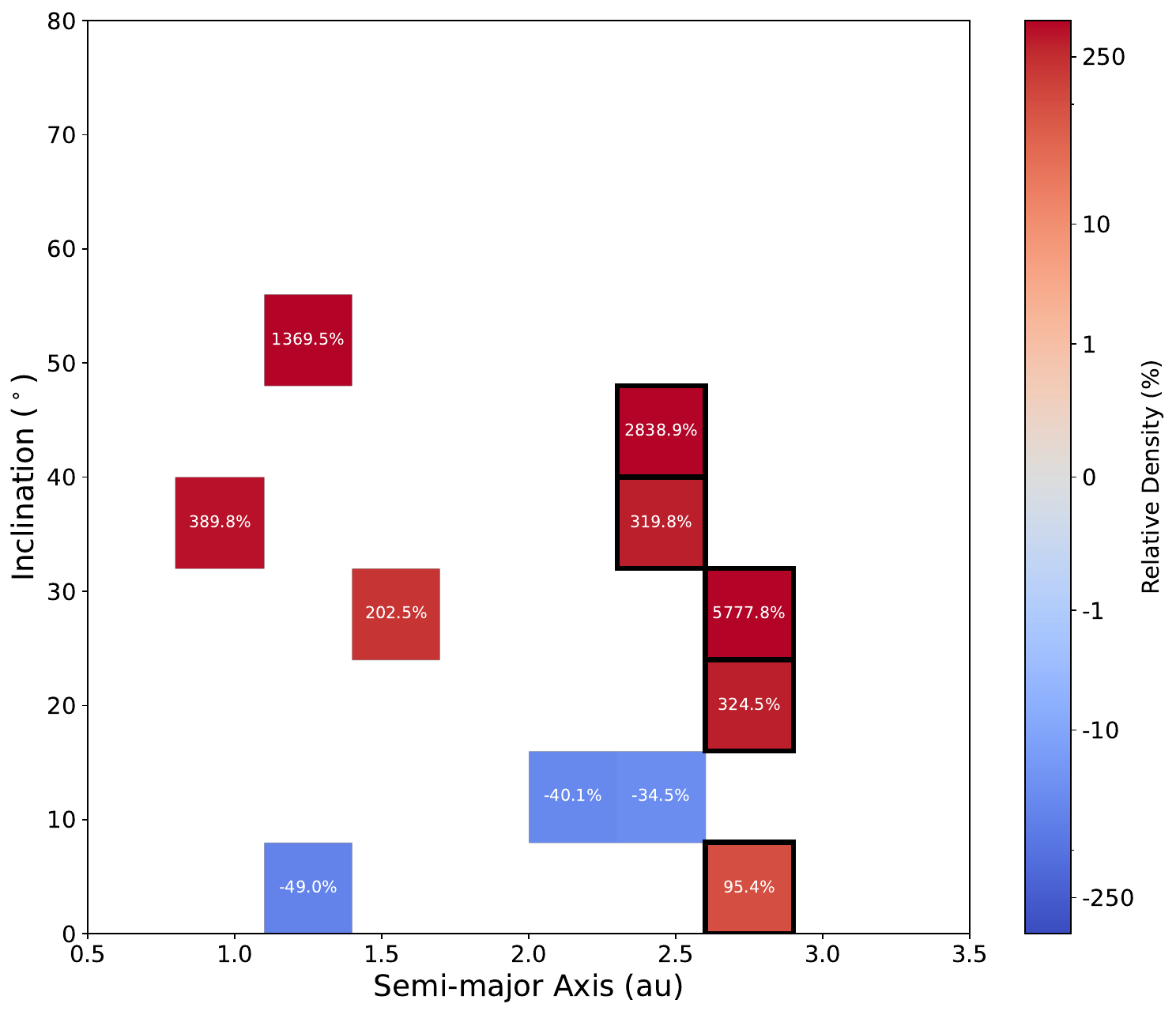} 
        \label{fig:image2}
    \end{subfigure}

    \vspace{-0.5cm} 

    \begin{subfigure}[b]{0.5\textwidth}
        \centering
        \includegraphics[width=\textwidth]{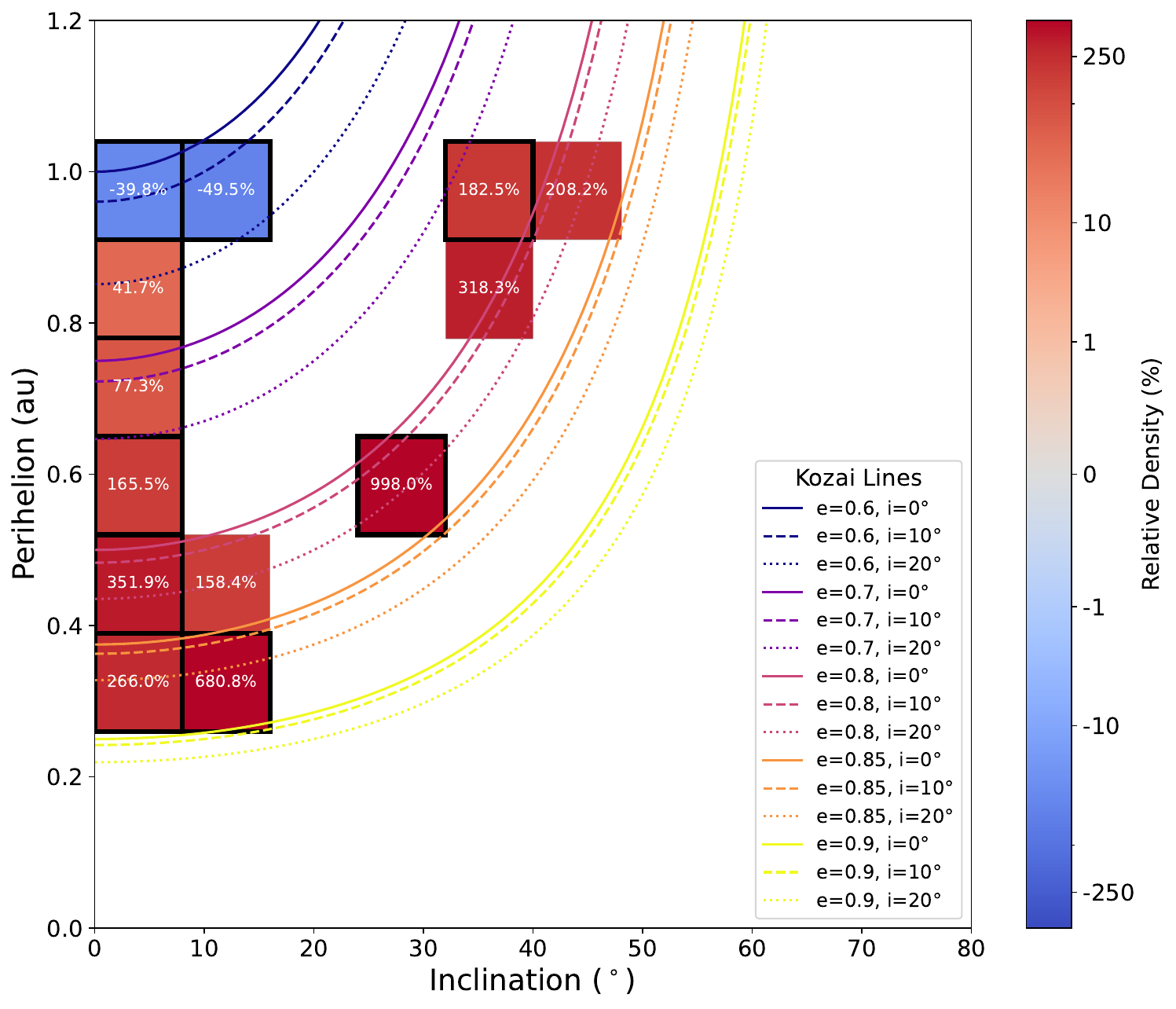} 
        \label{fig:image3}
    \end{subfigure}
    \hspace{-0.1\textwidth} 
    \begin{subfigure}[b]{0.5\textwidth}
        \centering
        \includegraphics[width=\textwidth]{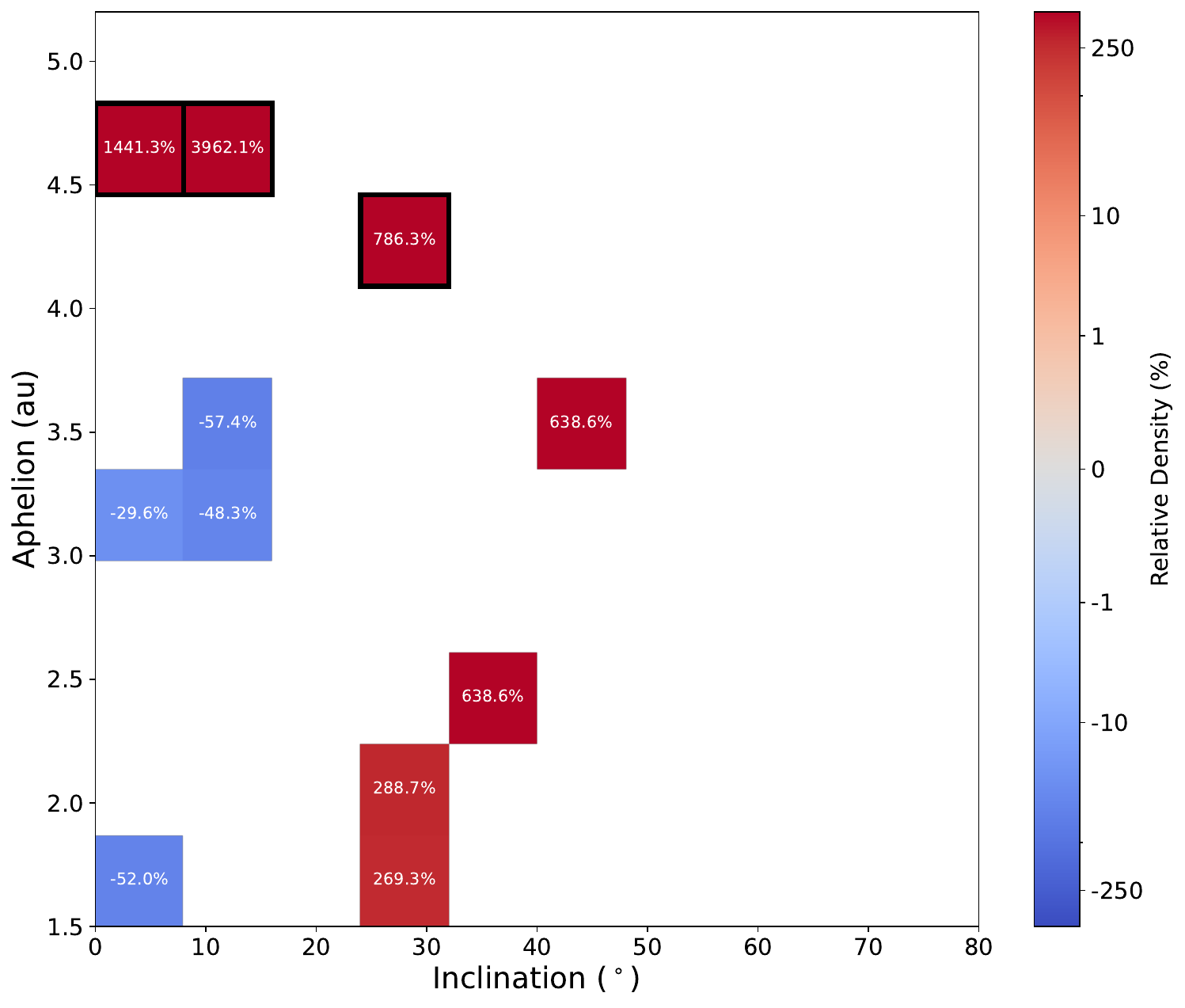} 
        \label{fig:image4}
    \end{subfigure}

    \caption{Orbital density heatmaps comparing sporadic $>10$\,g top-of-atmosphere asteroidal impacts and $>1$\,g meteorite falls. Panels show: (a) semi-major axis vs eccentricity, (b) semi-major axis vs inclination, (c) inclination vs perihelion distance, and (d) inclination vs aphelion distance. The colours indicate percentage differences in normalized density (red = overrepresented in the meteorite population; blue = underrepresented in the meteorite population). Statistical significance (chi-squared test) is highlighted by black bins at the 3$\sigma$ level and all others at 2$\sigma$. The dashed lines mark the perihelion evolution ($q$) as a function of inclination ($i$) under Kozai-Lidov resonance for orbits with $a=2.5$\,au.}
    \label{fig:differences}
\end{figure}

\subsection{The meteorite-dropping population}\label{sec2}
Next, we needed to identify a large enough meteorite fall dataset to draw statistically significant results from the differences between the two populations' orbital distributions. The GFO, EFN, and FRIPON fireball networks collected the fall database used here. In total, 540 possible $>$1\,g, 118 possible $>$50\,g, and 57 possible $>$1\,kg meteorite falls were identified between these three continental-scale fireball observation networks, which span 26 countries across regions in 5 continents, with a total collection area of approximately 12 million square kilometres \citep{colas2020fripon,devillepoix2020global,borovivcka2022_one}. 

Identifying meteorite-dropping fireballs is not a trivial problem and has been a continuously pressing issue within meteor science. Since the establishment of all-sky
fireball networks in the 1960s, the identification was difficult due to the lack of diverse observations of meteorite-dropping events. This was and is still somewhat particularly true for the low mass/energy events where only very small fragments were potentially produced on the ground. High-sensitivity cameras and automated detection algorithms have expanded the observational capabilities to capture events with lower luminosity and smaller fragment sizes \cite{jenniskens2011cams,vida2021global}. Additionally, the integration of Doppler weather radar data and advanced searching techniques using drones has enhanced our meteorite recovery rates \citep{fries2010doppler,Anderson2022ApJ}. Technological and methodological advancements have enhanced our ability to identify and recover meteorite-dropping fireballs. As a result, the meteoritic community can access a more representative sample of the solar system's material, advancing our understanding of planetary formation and evolution.

Based on fireball observations, the final mass estimates of meteorite fragments were determined using two primary methodologies. For the EFN data, we used mass estimates from \citet{borovivcka2022_one}, which were estimated using their fragmentation/ablation model. For the FRIPON and GFO networks, we employed the $\alpha$-$\beta$ methodology described by \citet{sansom2019determining}, which calculates the ballistic parameter and mass loss coefficient to identify fireballs likely to have dropped meteorites. This approach was used to estimate 1-gram, 50-gram, and 1\,kg meteorite falls within the FRIPON and GFO databases. 

An unexpected trend emerged when examining the impact of atmospheric filtering on the meteorite-dropping population. It is well established that velocity significantly influences the likelihood of meteoroid survival during atmospheric entry, with higher velocities generally associated with increased fragmentation and ablation \cite{brown2013meteorites}. Contrary to this expectation, our analysis revealed a bias toward higher velocities within the meteorite-dropping population (Fig.~2). This finding does not imply that higher-velocity objects have a better chance of surviving atmospheric entry—numerous studies have demonstrated the opposite. One might initially consider this result a potential artefact of the methods used to estimate meteorite-dropping fireballs; however, even when the minimum final mass threshold is increased to 1 kilogram -- where the likelihood of meteorite survival is almost certain -- the overrepresentation of higher velocities persists. The networks providing the top-of-the-atmosphere distributions all show this bias individually, though some networks to a lesser extent than the combined distribution. This consistency across different mass thresholds and datasets reinforces the robustness of the observation. Meteorite-dropping fireballs have a velocity distribution slightly shifted towards higher velocities relative to the top-of-the-atmosphere impact population, even though the atmosphere preferentially removes higher-velocity impactors.

\subsection{Atmospheric selection effects on the meteoritic record}\label{sec3}
The overrepresentation of high-velocity meteorite falls ($>$20\,km\,s$^{-1}$) suggests that the atmospheric filter alone does not fully account for the observed distribution at the top of the atmosphere. This observation indicates that something has preferentially modified the strengths of the population even before impact, resulting in a very durable higher-velocity component of the fireball population.

To explain the unexpected velocity variation between the top-of-the-atmosphere and the meteorite-dropping populations, two primary mechanisms could be responsible: (1) the collisional lifetimes of meteoroids or (2) thermal cycling-induced cracking and disintegration. The collisional lifetime refers to the average time a meteoroid can survive before being destroyed by a significant collision \citep{bottke1994collisional,gladman1997destination}, and it is necessarily linked to material strength, preferentially removing weaker meteoroids. Collisional lifetimes are also particularly shorter for meteoroids and asteroids that cross the main asteroid belt \citep{bottke1994collisional}, where they encounter more frequent collisions at high relative velocities. If collisional lifetimes were the dominant mechanism driving the higher-velocity component in the meteorite-dropping population, we would expect a significant drop-off in the relative occurrence of meteorite-dropping events when the orbit no longer crosses the main belt. This is not observed, as many recovered meteorite falls originate from such main-belt detached orbits (\url{https://www.meteoriteorbits.info/}).

On the other hand, thermal cracking and super-catastrophic disruption, driven by thermal effects and possibly micrometeoroid collisional grinding at low perihelion distances \citep{delbo2014thermal,granvik2016super,wiegert2020supercatastrophic}, provide a clearer link to the unexpected meteorite fall velocities. This mechanism is acting on a large swath of the population, with previous studies estimating that 80\% of NEOs reach below 0.05\,au at some point during their evolution \citep{granvik2018debiased,toliou2021minimum}. At low perihelion distances, small bodies experience repeated thermal stresses, causing fractures and weakening. Such thermal effects are well-documented in both observational studies (e.g., on Ryugu, Bennu, Didymos; \citealp{molaro2020situ,lucchetti2024fast}) and laboratory experiments \citep{delbo2014thermal,hazeli2018origins,libourel2021network}, where the cyclic heating and cooling near perihelion leads to structural breakdown of meteoritic material and the regolith formation on asteroidal surfaces. Additionally, the observed spectral slope decreases of S- and Q-type NEAs with perihelion distance have suggested that thermal fracturing could be the dominant mechanism at the origin of the rejuvenation of NEO surfaces \citep{graves2019resurfacing}.


As depicted in Figure~3, the perihelion distance of impacting meteoroids strongly correlates with their survival probability. Meteoroids with low perihelion distances (q\,$<$\,1\,au) are overrepresented in the meteorite population, while many objects impacting near q\,$\sim$\,1\,au fail to survive atmospheric entry. The relative excess in the meteorite fall population increases to nearly 5$\times$ that of the top of the atmosphere for low-inclination low-perihelion orbits (Fig.~3). This correlation suggests that perihelion reduction significantly shapes the meteoroid population that ultimately reaches Earth. Resonant mechanisms, such as mean-motion and secular resonances, are the primary drivers that reduce the perihelion distance of asteroids. Previous studies have shown that mean-motion resonances (MMRs) with Jupiter, such as the 3:1 MMR and 4:1 MMR, are particularly effective at driving the eccentricities of near-Earth asteroids (NEAs) to high values \citep{toliou2023resonant}. Similarly, secular resonances, like $\nu_6$ and $\nu_5$, contribute to this reduction by inducing oscillations in the eccentricities of asteroids, further pushing their perihelia inward \citep{toliou2023resonant}. These resonances can systematically lower the perihelion distance over long timescales, with the $\nu_6$ resonance being particularly influential for bodies originating from the inner main asteroid belt. At the same time, $\nu_5$ plays a more auxiliary role by affecting asteroids at lower semi-major axes. These resonances drive asteroids toward smaller perihelia and increase their eccentricity, making them more likely to experience frequent close approaches to the Sun and, consequently, undergo thermal and collisional processing that weakens them before atmospheric entry.

Moreover, Kozai-Lidov resonance, especially for objects on inclined orbits, can cyclically exchange inclination for eccentricity, lowering the perihelion distance over time. This resonance can trap objects in periodic close solar approaches, contributing to the observed overabundance of low-perihelion meteoroids in the meteorite population \citep{wiegert2020supercatastrophic,toliou2023resonant,shoberjfcs2024}. Simulations indicate that some objects may experience large oscillations in eccentricity due to the Kozai-Lidov effect, which, when combined with MMRs, can further accelerate the inward migration of their orbits \citep{toliou2023resonant}. Thermal cycling, combined with resonant interactions, progressively eliminates fragile and hydrated objects, favouring the survival of high-strength stony materials that dominate the meteorite record. This observation of a perihelion and atmospheric filter removing the weakest meteoroids from the population thus supports the claim the discrepancy between the expected carbonaceous meteorite flux ($\sim$50\%) and the observed one ($\sim$4\%) \citep{brovz2024source}. We find that for impactors $>$1\,kg (based on EFN, GFO, and FRIPON observations), the survival rate of a 50\,g meteorite is roughly around 30-50\%, also tentatively consistent with the mismatch in the carbonaceous meteorite production rate. Additionally, the CI chondrites are expected to have a flux that is 2$\times$ what is observed relative to other carbonaceous chondrites; however, the families proposed as sources for CIs (Polana, Euphrosyne, Clarissa, Misa,
Hoffmeister) have lower inclinations and smaller semi-major axes compared to the proposed source families of CMs (e.g., Veritas, K\"{o}nig, etc.). These lower on average semi-major axis values of the CI source families are more amenable to the transport to low-perihelion values, possibly removing them more quickly from thermal cracking \citep{granvik2018debiased,toliou2021minimum,toliou2023resonant}. We do observe an excess of weak meteoroids on slightly inclined orbits (10-20$^{\circ}$) with semi-major axis values between $\sim$\,2.0-3\,au(Fig.~3), corresponding to the location of the K\"{o}nig family, possibly hinting it is the cause. The significance does achieve a 2$\sigma$ level for two bins. This excess of weak meteoroids is consistent with the K\"{o}nig family and potentially also a Polana source at low inclination, but the results vary depending on the minimal meteorite masses used. Thus, more observations are needed to confirm. The survival trend that persists no matter the size range, dataset, or minimum deceleration used while still achieving at least 2$\sigma$ significance is associated with perihelion distance. 

Thermal stresses explain the excess of robust meteoroids at perihelion distances below 1\,au and their increased ability to survive atmospheric entry. However, another marked aspect is the concentration of meteoroids with perihelion distances near 1\,au, which tend to be weaker than their low-perihelion counterparts. Two factors are responsible: (1) these meteoroids have not undergone significant perihelion lowering in their past \citep{toliou2021minimum}, and (2) the excess of the population, which is likely linked to tidal disruptions (e.g., \citealp{schunova2014properties,jopek2020orbital,granvik2024tidal,shober2024decoherence}), is more likely to produce inherently weaker debris because carbonaceous bodies are more amenable to such disruptions. 

\section{Discussion}\label{discussion}
Based on nearly 8,000 impacts observed in our atmosphere, it is clear that the atmospheric filter is not the entire story. An unambiguous modification and selection of meteoroids occurs even before an impact with the atmosphere. We find that perihelion filtering, likely resulting from excessive thermal stresses, is the primary process that reduces the hydrated carbonaceous material to smaller size ranges \citep{wiegert2020supercatastrophic,brovz2024source}. High-strength meteoroids dominate the meteorite fall record, having survived repeated closer approaches to the Sun, where weaker hydrated materials would have been eroded or fragmented. This perihelion filtering effect plays a significant role in shaping the meteorite population on Earth, as these high-strength meteoroids appear more resilient to atmospheric entry and are more highly represented in our samples.


On the other hand, our analysis also suggests that meteoroids formed through tidal disruptions are generally weaker and less likely to survive atmospheric entry compared to the average impactor. A clustering analysis of 7,982 impacts at the top of the atmosphere identified around 0.4\% to be linked with NEA clusters (see Methods). In contrast, only about 0.2\% of the 540 possible 1\,g falls could be linked to tidally created NEA/top-of-the-atmosphere clusters \citep{shober2024decoherence}. This implies that meteoroids associated with tidally created streams are approximately twice as likely to fail atmospheric entry due to their anomalously weak structure. Numerous studies have explored meteoroid streams as a potential consistent impact threat to Earth \citep{halliday1987detection,pauls2005decoherence,koten2014search}, but tidal disruption debris appears to be significantly weaker for centimetre- to metre-sized objects. Despite their increased impact probabilities, these meteoroids contribute minimally to the meteorite population. Furthermore, the estimated percentage of clustered NEAs is around 0.13\% \citep{jopek2020orbital,shober2024decoherence}, lower than what we observe for the top-of-the-atmosphere population. The centimetre- to metre-sized population contains more tidally-generated debris relative to NEOs. This is consistent with models of tidally-disrupted NEA families, which predict a steeper size-frequency distribution compared to the general NEO population \citep{schunova2014properties}.

This dramatic increase in friability is more compatible with carbonaceous-dominated compositions. Ordinary chondrite meteorite falls tend to reach peak dynamical pressures of 3-5 MPa \citep{borovivcka2020two}; however, carbonaceous falls disrupt at much lower pressures (0.5\,MPa; \citealp{borovivcka2019maribo,mcmullan2024winchcombe}) while traversing the atmosphere. Thus, the low survivability of tidally-disrupted meteoroids could be explained by a carbonaceous composition and the specific impact conditions (e.g., low-impact angle, lower relative velocity, etc.) needed to persist \citep{borovivcka2019maribo,mcmullan2024winchcombe,brovz2024source}. Moreover, the presence of volatiles within these carbonaceous materials can significantly reduce their mechanical strength, contributing to both higher degrees of porosity (2-3$\times$ higher for CI/CMs; \citealp{flynn2018physical}) and an increased tendency to fragment upon atmospheric entry \citep{trigo2009tensile}. This is also supported by the abundance of carbonaceous compositions and mineralogical features amongst micrometeorites \citep{genge2020micrometeorites}. Such differences in porosity and strength among meteorite classes reflect their distinct collisional compaction histories \citep{blum2006physics,trigo2009tensile}, with carbonaceous chondrites typically retaining higher bulk porosities and lower bulk densities than ordinary chondrites \citep{flynn2018physical}. Moreover, recent evidence based on the irradiation history, isotopic geochemistry, and structural heterogeneity of CI/CM carbonaceous meteorites suggests that some CI/CM samples are more consistent with a recent ejection in near-Earth space \citep{shober2024cosmic}, possibly tying it to tidal disruptions or meteoroid impacts. This recent lithification and release in near-Earth space provides a more compelling explanation for some of the salient features found in CI/CM chondrites \citep{shober2024cosmic}. This contrasts with traditional models assuming long exposure in the main asteroid belt, where the mixing and particle flux conditions would be less favourable for such a pattern. Recent observations of fluid flow and regolith activity in NEAs further support the idea that these processes are ongoing and that the lithification of CI/CM chondrites may be more recent than previously thought \citep{turner2021carbonaceous,shober2024cosmic}. 

In-situ observations of C-type rubble-pile asteroids Ryugu and Bennu lend further insight into this discrepancy. Both Ryugu and Bennu are known to be composed of highly porous boulders with low tensile strength and high bulk porosity \citep{grott2019low,rozitis2020asteroid}, and analysis of returned samples supports these remote observations \citet{yada2022ryugu,lauretta2024asteroid}. Yet, it was also found that the fine regolith production on Bennu is frustrated in areas where these porous boulders are more concentrated \citep{cambioni2021fine}, implying that high-porosity carbonaceous material is more resilient to thermal cycling and its associated stresses. While porous boulders on Bennu or Ryugu may survive thermal cycling for longer periods, they would certainly completely disintegrate during an atmospheric passage. The weakness of carbonaceous meteoroids in near-Earth space could be related to porosity and its relation to thermal resilience. The perihelion filter eliminates more compact, lower porosity, lower-perihelion carbonaceous meteoroids due to thermal cycling, and the atmosphere filter eliminates those with high porosity, leaving only the lucky, compact, strong material in the meteorite collections that happened to reach the Earth quickly. In other words, compact, tidally-generated debris is favoured in this scenario. 

Despite the concentration of weaker meteoroids near q\,$\sim$\,1\,au and $\iota\sim$\,0$^{\circ}$, of the 50+ meteorites with precisely measured orbits, H and L chondrites dominate in this region, making up about 75\% of the 19 falls in this region. This is, however, consistent with what one would expect, as the weaker meteoroids are not surviving the atmospheric passage. Albeit, a lack of observation does not constitute one. Thus, further work would need to be done to definitively associate the weaker meteoroids observed in this region with carbonaceous material. Interestingly, the Almahata Sitta meteorite (asteroid 2008 TC3) also originated from an orbit with q\,$\sim$\,1\,au, and low inclination \citep{jenniskens20042003}. This meteorite is noteworthy for its polymict breccia composition, containing multiple meteorite types within a single parent body. While not consistent with being a rubble pile (exhibiting flares at 0.3 – 1.3\,MPa, i.e., much higher than 25 Pa
expected for rubble piles; \citealp{borovivcka2009meteosat}), asteroid 2008 TC3 was very likely derived from one. This could indicate that a broader rubble-pile boulder weakness could also contribute to these meteoroids' observed friability. In-situ observations of S-type asteroids, such as Eros, Itokawa, and Didymos, also reveal evidence of higher bulk porosities, but with thermal inertia values more than two times greater than those measured on Bennu and Ryugu. This suggests that while the boulders on these S-type asteroids might be more likely to survive the atmosphere, they experience higher levels of thermal fatigue and cracking. This higher tendency to form cracks is also consistent with the pressures experienced during fragmentation of ordinary chondrite meteorite falls \citep{popova2011very,borovivcka2020two}. While thermal effects are a key factor, mechanical stresses induced by collisions between asteroids or meteoroids can also produce cracks \citep{borovivcka2020two}, further weakening the meteoroids and increasing their likelihood of disruption. The thermal cracking and fatigue observed in rubble-pile asteroids and its potential to contribute to the weakness of meteoroids that fail to survive atmospheric entry remains a topic that warrants further investigation. Advances in fireball observations, particularly in high-resolution photometry and spectral analysis, combined with better-calibrated fragmentation models, hold the potential to significantly improve our ability to identify and characterize the compositions of meteoroids. With these tools, we may be able to better differentiate carbonaceous meteoroids from other types, even before they enter the atmosphere \citep{borovivcka2020two,loehle2024meteorite,vaubaillon2024uv,matlovivc2024spectral}, allowing for a deeper understanding of their origins and evolution. 

\section{Methods}\label{sec11}

\subsection{Data collection}
The data used in this paper is sourced from six different systems, encompassing 19 observation networks. These include the European Fireball Network (EFN), European viDeo Meteor Observation Network Database (EDMOND), Cameras for Allsky Meteor Surveillance (CAMS), the Global Meteor Network (GMN), Fireball Recovery and InterPlanetary Observation Network (FRIPON), and the Global Fireball Observatory (GFO). Notably, EDMOND is not an observation network but rather a comprehensive database that compiles data from various European-based video meteor observation networks, making the information publicly accessible. In contrast, the other sources (EFN, CAMS, GMN, FRIPON, and GFO) are all networks or network collaborations that directly capture meteor observations. Statistical significance was achieved in our analysis only by including data from such a broad array of networks. EDMOND, CAMS, and GMN focus more on detecting smaller meteoroids, mostly milligrams to grams in mass. Meanwhile, EFN, FRIPON, and GFO are tailored to larger meteorite-dropping events. For the top-of-the-atmosphere population, comprising 7,982 events, data was drawn from EDMOND, CAMS, GMN, FRIPON, and EFN. The GFO was excluded from this population due to biases related to its operational limitations, which are more challenging to correct (see discussion of \citep{shoberjfcs2024}). The meteorite-dropping population focused on events with at least 1-gram meteorite potential, totalling 540 events sourced from FRIPON, EFN, and GFO, as these networks prioritize observing and recovering meteorite falls. 

\subsubsection{EFN}
The European Fireball Network (EFN), founded in 1963, has undergone significant technological advancements, continually enhancing its meteor detection capabilities over the decades \citep{borovivcka2022_one,borovivcka2022_two}. Throughout its history, the EFN has contributed to the recovery of at least 13 meteorites \citep{spurny2003photographic,spurny2017ejby,borovivcka2019maribo,spurny2020vzvdar}. The network primarily utilizes digital autonomous fireball observers (DAFOs), fully automated, weatherproof systems that continuously monitor the sky under favourable conditions. These systems employ DSLR cameras with fisheye lenses, similar to those used by the GFO, to capture detailed images. EFN’s cameras are designed to detect meteors brighter than magnitude -2, and they can perform radiometric measurements on meteors brighter than magnitude -4 \citep{borovivcka2022_one}. The network can detect meteoroids larger than 5 grams and record high-velocity meteoroids with masses as low as 0.1 grams. EFN comprises 26 stations across Central Europe, collectively covering about one million square kilometres.

For this study, 824 EFN fireballs were sourced from previous publications \citep{borovivcka2022_one,borovivcka2022_two}. The network utilizes the straight-line least squares (SLLS) method to calculate the atmospheric trajectories of fireballs, assuming a linear space trajectory, similar to the DFN \citep{Borovicka_1990BAICz}. However, EFN's velocity calculation method differs, relying on a four-parameter model that incorporates pre-atmospheric velocity, ablation coefficients, and mass-related factors \citep{pecina1983BAICz..34..102P}. Atmospheric models such as CIRA72 or NRLMSISE-00 are used to fine-tune the trajectory, and when necessary, manual adjustments are made for significant deceleration. Systematic discrepancies between cameras are resolved before the final data analysis. Heliocentric orbits are then calculated using a slightly modified version of the method developed by \citet{Ceplecha_1987}, accounting for Earth's gravitational pull and rotation.

\subsubsection{EDMOND}
The European viDeo Meteor Observation Network Database (EDMOND) is a collaborative international effort involving several European meteor networks. The database is a significant resource for recording and analyzing meteor activity across Europe. Established through cooperation among various national networks, EDMOND integrates data from organizations such as BOAM (Base des Observateurs Amateurs de Météores) in France, CEMeNt (Central European Meteor Network), HMN (Hungarian Meteor Network), IMTN (Italian Meteor and TLE Networks), PFN (Polish Fireball Network), SVMN (Slovak Video Meteor Network), and UKMON (UK Meteor Observation Network). Observers from Bosnia, Serbia, Ukraine, and other countries also contribute to the database, which has grown significantly since its inception \citep{kornos2014edmond}. EDMOND collects data using two primary software systems: MetRec and UFO Tools. MetRec, created by Sirko Molau, is widely used in Central Europe, while the UFO Tools system developed by SonotaCo is popular in other networks \citep{sonotaco2009meteor}. Both tools capture single-station meteor observations, and the data are compiled into the EDMOND database using UFOOrbit software. This software processes and filters the data to produce meteor orbits, applying specific criteria such as velocity limits and trajectory angles to ensure accuracy \citep{kornos2014edmond}.

The database includes over 7.4 million individual meteor observations, resulting in more than 480,000 meteor orbits. These observations span from 2000 to 2023 and are freely accessible online (\url{https://www.meteornews.net/edmond/edmond/edmond-database/}).

\subsubsection{CAMS}
The Cameras for Allsky Meteor Surveillance (CAMS) is an innovative network established to verify minor meteor showers by tracking meteoroid orbits using multi-station video observations. Launched in 2010, CAMS combines low-light video technology with advanced data processing methods to measure meteoroid trajectories with high precision. The primary goal of the CAMS project is to validate and discover new meteor showers, especially those associated with newly identified NEOs. With over 560 video cameras deployed across 9 countries, CAMS covers the sky above 31° elevation, detecting meteors as faint as magnitude +4 \citep{jenniskens2011cams}.

The CAMS network captures simultaneous meteor detections from multiple stations, allowing for accurate triangulation of their atmospheric trajectories. Each station is equipped with Watec 902 H2 Ultimate low-light video cameras, offering a high level of sensitivity with a stellar limiting magnitude of +5.4. Each CAMS station typically operates with 20 cameras. The cameras are connected to servers that record video overnight, which is subsequently processed during the day to detect linear streaks indicative of meteors. The detected meteors are processed using a coincidence algorithm, which identifies meteors observed from different stations and calculates their atmospheric trajectory and velocity using the Borovička method \citep{jenniskens2016established}. The software accounts for deceleration due to atmospheric drag and gravitational influences to determine the pre-atmospheric velocity and trajectory of the meteoroids. 

CAMS data are submitted regularly to the IAU Meteor Data Center and made publicly available on the CAMS website (\url{http://cams.seti.org/}). The data used in this study was taken from a subset of the 471,582 total meteors available at the IAU Meteor Data Center (\url{https://ceres.ta3.sk/iaumdcdb/}).  

\subsubsection{GMN}
The Global Meteor Network (GMN; \url{https://globalmeteornetwork.org/}) is an international initiative designed to conduct continuous optical observations of meteors using a distributed network of low-cost CMOS video cameras and Raspberry Pi computers. Since its launch in 2018, the GMN has rapidly expanded, now comprising over 450 cameras across 30 countries. The primary goal of the network is to offer long-term monitoring of meteor showers, meteoroid flux, and size distribution within the optical meteor mass range. Additionally, it seeks to enhance public awareness of the near-Earth meteoroid environment by publishing orbital data within 24 hours of detection, promoting transparency and reproducibility in data reduction methods \citep{vida2021global}.

GMN systems have wide-field cameras that capture a stellar limiting magnitude of +6.0$\pm$0.5, operating at 25 frames per second. These systems have accumulated over 220,000 precise meteoroid orbits from December 2018 to mid-2021, achieving a median radiant precision of 0.47$^{\circ}$, with approximately 20\% of meteors observed from four or more stations achieving a precision of 0.32$^{\circ}$. Such precision is crucial for accurately measuring the physical dispersions of meteor showers \citep{vida2021global}.

GMN's design philosophy emphasizes cost-efficiency, openness, and decentralized participation. Most systems consist of low-light Internet Protocol (IP) cameras and accessories costing around \$200 USD per station, utilizing open-source software called Raspberry Pi Meteor Station (RMS) to automatically detect and analyse meteors. The system performs frequent recalibrations to ensure accuracy, applying novel astrometric and photometric methods to adjust for potential environmental and instrumental variations. The GMN’s software framework is designed to ensure reproducibility and transparency, with data and observations made publicly accessible \citep{vida2021global}. The GMN's automated systems detect, process, and upload meteor observations to a central server, making the data available within hours. 

\subsubsection{FRIPON}
Fireball Recovery and InterPlanetary Observation Network (FRIPON; \url{https://www.fripon.org/}) is an international project launched to improve the tracking and recovery of meteoroids and meteorites, contributing to the understanding of interplanetary material and solar system evolution. Initiated in France in 2015, the network quickly expanded to include neighbouring countries, now covering over 2 million square kilometres across 15 nations on four continents, with over 250 cameras and 40 radio receivers \citep{colas2020fripon}. The network's cameras, which are wide-angle CCDs, capture images at 30 frames per second, providing high temporal resolution data. Longer exposures are taken every 10 minutes to enhance astrometry and photometry, improving the signal-to-noise ratio up to a magnitude of 6 at dark sites \citep{anghel2019photometric,jeanne2019calibration,jeanne2020methode,colas2020fripon}. FRIPON’s automation supports prompt meteorite recovery efforts, targeting meteorites with final masses of 500 grams or larger. The network has recovered seven meteorites so far, aided by its expansive international collaboration.

The standard data reduction process for FRIPON is documented in several studies \citep{jeanne2019calibration,jeanne2020methode,colas2020fripon}. However, this research processed FRIPON observations using a Monte Carlo triangulation method \citep{vida2020estimating}. The method begins with an initial trajectory based on the intersection of observed meteor paths, with angular residuals used to gauge uncertainty \citep{Gural_2012M&PS,Weryk_Brown_2012P&SS}. These uncertainties guide the generation of Monte Carlo simulations, which introduce Gaussian noise to observations and yield a range of plausible trajectory solutions. A key innovation is selecting the most consistent solution based on meteor dynamics across multiple observation stations. The open-source Python code used in this process (\url{https://github.com/wmpg/WesternMeteorPyLib}) ensures the transparency and reproducibility of the method.

\subsubsection{GFO}
The Desert Fireball Network (DFN) is an expansive array of cameras spanning approximately 2.5 million square kilometres of Australia's outback, covering over one-third of the country’s land area, making it the most extensive fireball monitoring system globally \citep{howie2017build}. Development began in 2013 with a focus on creating reliable, autonomous systems, culminating in the deployment of the final design between 2014 and 2015, which included around 50 fireball stations \citep{howie2017build}. The DFN’s desert locations are chosen to increase the chances of successful meteorite recoveries. These stations use high-resolution DSLR cameras paired with all-sky fisheye lenses, utilizing long-exposure photography enhanced by GNSS-synchronized liquid crystal shutters. This configuration allows for precise meteorite fall predictions, even under conditions with limiting magnitudes of zero. The DFN is capable of detecting meteoroids as small as 5 centimetres before deceleration becomes significant and capturing brighter phases of larger meteoroids despite some sensor limitations \citep{howie2017submillisecond,devillepoix2019observation}. To determine the atmospheric trajectories of fireballs, the DFN employs a straight-line least squares (SLLS) method and velocity profiling through an extended Kalman smoother \citep{Borovicka_1990BAICz,sansom2015novel}. This process accounts for observational uncertainties, and the Kalman filter enhances the accuracy of the data. For pre-entry orbit calculations, meteoroid states are integrated beyond Earth's gravitational influence, considering major perturbative forces.

The DFN has expanded globally, forming the Global Fireball Observatory (GFO) in collaboration with 18 institutions across nine countries all using the same observation technology developed by the DFN \citep{devillepoix2020global}. The GFO has thus far been instrumental in recovering 15 meteorites, representing roughly 30\% of all recoveries with associated orbital data \citep{King_Winchcombe2022SciA,shober2022arpu,Devillepoix_Madura_Cave,Anderson2022ApJ}. 

\subsection{Mass Estimations}
Mass estimations for meteoroids based on meteor and fireball observations are generally classified into two broad categories: photometric and dynamic. Photometric mass estimations rely on the observed brightness of the meteor or fireball and the conversion of kinetic energy into visible light, known as luminous efficiency. This method estimates mass based on the total emitted light. On the other hand, dynamic mass estimations utilize the meteoroid's deceleration and velocity profile as it passes through the atmosphere, estimating mass based on its interaction with atmospheric drag. This method often requires assumptions about the meteoroid’s physical parameters, such as shape, density, and ablation properties. 

In this study, assessing how much the orbital distribution varies as a function of mass is essential. If strong correlations between mass and orbital variation are observed within the chosen size range, adjustments would need to be made to account for this difference. However, by focusing on a mass range where orbital variation as a function of mass is minimal, we can avoid weighting the orbital distribution based on the size-frequency distribution. This allows us to simplify the analysis while ensuring that the observed orbital differences between the top and bottom of the atmosphere are not biased by mass variations. 

\subsubsection{EFN}
The initial mass of the meteoroids observed by the EFN is estimated using the photometric method, which assumes that the radiated energy is proportional to the loss of kinetic energy. The method is described, and the data is publicly available in \citet{borovivcka2022_one}. The meteoroid's initial mass is computed as:

\begin{equation}
    m_{\text{phot}} = \int  \frac{2}{\tau(v) v^2} I(t) \, dt
    \label{eq:photo_mass}
\end{equation}

where \( v \) is the velocity of the meteoroid, \( \tau(v) \) is the luminous efficiency, \( I(t) \) is the radiated energy, and \( t \) is time. The radiated energy \( I(t) \) is derived from the absolute magnitude \( M \) using the formula:
\(I = I_0 \cdot 10^{-0.4M}\) where \( I_0 = 1500 \, \text{W} \) is the energy of a zero-magnitude meteor, following the method described by \citet{ceplecha1998meteor}.

The luminous efficiency \( \tau(v) \), representing the fraction of kinetic energy converted into light, is a function of velocity and is expressed as:

\begin{equation}
    \ln \tau = 
    \begin{cases} 
    0.465 - 10.307 \ln v + 9.781 (\ln v)^2 - 3.0414 (\ln v)^3 \\ + 0.3213 (\ln v)^4, & \text{if } v < 25.372 \, \text{km/s} \\
    -1.53 + \ln v, & \text{if } v \geq 25.372 \, \text{km/s}
    \end{cases}
    \label{eq:luminous_effeciency}
\end{equation}

where \( v \) is the velocity in km/s. This equation, from \citet{revelle_ceplecha_2001}, provides the luminous efficiency for a meteoroid with an approximate mass of 10 kg. Additionally, for smaller meteoroids (less than 1 kg), the luminous efficiency may be about two times lower than that of larger ones.

The mass obtained from the light curve is referred to as the photometric mass \( m_{\text{phot}} \). A dynamic mass can also be estimated from the meteoroid's deceleration and velocity measurements, but it assumes no fragmentation during atmospheric entry. Therefore, the photometric mass is often considered more reliable in practical cases, even though the luminous efficiency parameter can vary by a couple of orders of magnitudes \citep{loehle2024meteorite}.

\subsubsection{EDMOND/CAMS}
For the CAMS and EDMOND datasets, which are both publicly accessible through the IAU’s Meteor Data Center(\url{https://ceres.ta3.sk/iaumdcdb/}), meteoroid masses were not directly reported. Therefore, we estimated the masses using the method based on \citet{verniani1973analysis} and \citet{hughes1995perseid}. The mass estimation is derived from the observed visual magnitudes and atmospheric velocities using the following equation in S.I.:

\begin{equation}
    \log m (\text{g}) = 14.7 - 4.0 \log V (\text{cm/s}) - 0.4 M_v
    \label{eq:hughes_mass}
\end{equation}

where \( m \) is the meteoroid mass in kilograms, \( V \) is the meteoroid's atmospheric velocity in m/s, and \( M_v \) is the observed visual magnitude of the meteor. This equation estimates the meteoroid mass by relating its velocity and visual magnitude to the kinetic energy converted into light during atmospheric entry. The equation assumes meteoroids ablate efficiently in the atmosphere, converting their kinetic energy into visible light. The term \( 14.7 \) is a constant derived from theoretical and empirical studies, and the \( -4.0 \log V \) term reflects the influence of velocity on brightness — faster meteoroids emit more light, requiring an adjustment in mass estimates. The term \( -0.4 M_v \) accounts for the observed visual magnitude, where brighter meteors (lower \( M_v \)) correspond to more massive objects.

The method assumes that the meteoroids do not undergo significant fragmentation, meaning their mass loss primarily results from ablation, as the single-body theory describes. This approach also assumes a standard value for the meteoroids' luminous efficiency (0.4-1\%), which is critical in converting observed magnitude to mass. The ablation coefficient and other parameters are derived from the ionization curve \citep{verniani1973analysis}, which correlates the maximum electron line density with the meteoroid's physical properties.

\subsubsection{GMN}
The GMN operates using an automated data reduction pipeline, a detailed description of which can be found in \citet{vida2021global}. The mass estimation for meteors observed by GMN is derived from integrating a 100 km range-corrected light curve. This method involves utilizing the bolometric power of a zero-magnitude meteor, \( P_{0m} = 1210 \, \text{W} \), which has been calibrated for specific camera sensors like the Sony HAD EX-View. A luminous efficiency value \( \tau \), set at 0.7\%, is applied in line with previous studies. To calculate the light curve, observations from multiple stations are combined, and the brightest observation is selected to minimize the effects of camera saturation. This is particularly relevant for meteors observed from more distant stations, which are less likely to saturate. GMN cameras typically saturate around magnitude -1. The final photometric mass estimate is then obtained from this process. In cases where only part of the meteor’s trajectory is observed (e.g., due to field-of-view limitations), the mass is flagged as a lower limit \citep{vida2021global}. 

\subsubsection{FRIPON}
The mass estimation for the fireballs detected by the FRIPON network was performed using a photometric approach.
Thus, the light curve in absolute magnitudes (referenced at a range of 100 km), was converted into radiated energy using
\begin{equation}
\centering
    \begin{aligned}
           I = I_{0} \cdot 10 ^{\left(\dfrac{M}{-2.5}\right)}
    \end{aligned}
\label{eq:intensity}
\end{equation}

from \cite{ceplecha1998meteor}, where I$_{0}$ is the radiative output of a meteor of zero absolute magnitude light source. Here the value of 1500 W was used, as others have done \cite{spurny2020vzvdar, spurny2017discovery, 1996A&A...311..329C}. Next, the radiated energy was converted into source energy via the correspondence relation derived by \citet{2021MNRAS.508.5716A} from well-known atmospheric impacts. Finally, the entry velocity obtained via dynamic modelling \citep{jeanne2019calibration}, along with the source energy, was plugged into the kinetic energy equation to obtain the pre-entry mass of the object.

The resulting nominal luminous efficiency was 0.367, with lower and upper limits of 0.222 and 0.606. This range aligns with previous studies on fireball networks and laboratory experiments, indicating that luminous efficiency for meteoroids can vary widely based on their size, composition, and velocity \citep{borovivcka2022_one, loehle2024meteorite}. Studies such as \citet{loehle2024meteorite} have shown that luminous efficiencies typically range between 0.01\% and 1\%, with higher values observed for materials with higher iron content. However, the radiation emission becomes more complicated for larger metre-scale bodies, and these values can get several times larger as well  \citep{nemtchinov1997assessment}. 

\subsubsection{GFO}
None of the GFO data was used in this study to estimate the top-of-the-atmosphere population's orbital distribution. This decision is due to the difficulty of removing some intrinsic observational biases associated with GFO sensors (discussed in the following section). Therefore, we solely utilize GFO data for its meteorite-dropping components. Importantly, the GFO typically employs an Extended Kalman Smoother (EKS) filter for estimating dynamic masses \citep{sansom2015novel}, which is particularly well-suited for uncertainty estimation in mass predictions. The EKS filter predicts changes to the meteoroid’s state, including its position, velocity, and mass, using the single-body aerodynamic equations \citep{sansom2015novel}. While the GFO stations do not currently have dedicated sensors for calibrated photometry, and photometry is not a standard output of the GFO pipeline, the EKS filter has been very accurate in predicting final masses for meteorite-dropping events. To date, GFO sensors have been involved in the successful recovery of 17 meteorite falls (\url{https://dfn.gfo.rocks/meteorites.html}). The dynamic mass estimates provided by the EKS filter are particularly suitable for meteorite-dropping fireballs, where the meteoroid undergoes significant deceleration, allowing for more precise dynamic mass estimations. However, the dynamic mass estimates become more uncertain for meteoroids burning up high in the atmosphere with minimal deceleration or those with significant fragmentations.

\subsection{Meteorite Fall Identification}
Identifying meteorite-dropping events is a crucial aspect of our analysis, enabling us to discern the potential meteorites from the larger population of observed fireballs. This study identified meteorite fall candidates using the EFN, FRIPON, and GFO datasets. 

For the EFN, terminal masses derivation is described in  \citet{borovivcka2022_one}. The meteoroid mass at the last measured velocity point is estimated using a physical four-parameter velocity model for the entire trajectory, assuming a shape-density coefficient of $\Gamma A = 0.7$ and a meteoroid density of $\rho_d = 3000$ kg m$^{-3}$. \citet{borovivcka2022_one} only provide an estimate if the mass is $\geq$1\,g and the terminal velocity is $<$10 km s$^{-1}$; otherwise, the mass is set to zero. 

For FRIPON and GFO, meteorite-dropping events were identified using the $\alpha-\beta$ methodology \citep{sansom2019determining}. The $\alpha-\beta$ criterion is a powerful tool for quickly assessing whether a fireball is likely to drop meteorites. This method calculates two key parameters: the ballistic coefficient ($\alpha$), which relates to the mass and drag properties of the meteoroid, and the mass-loss parameter ($\beta$), which characterizes how much mass the meteoroid loses during its descent. Using these parameters, the $\alpha-\beta$ diagram allows for the separation of meteorite-dropping events from the general population based on their physical deceleration behaviour.

In this study, we applied the $\alpha-\beta$ criterion to FRIPON and GFO data using three minimum final mass limits: 1 gram, 50 grams, and 1 kilogram. The limiting values to be considered a potential meteorite-dropping event were calculated using a carbonaceous chondrite bulk density of 2400\,kg\,m$^{-3}$, to avoid an over-estimation of the filtering of carbonaceous debris. The $\alpha-\beta$ methodology offers a robust framework for predicting meteorite falls, with previous studies demonstrating its effectiveness in identifying viable meteorite recovery targets. Only meteoroid impacts that were observed to undergo at minimum 20\% deceleration during the bright-flight phase were included in the study, as the $\alpha-\beta$ estimation is estimated based on the deceleration profile. 

The equations used to describe the regions below where, given a shape change coefficient ($\mu$), a meteorite of at least 50\,g or 1\,kg would survive are laid out in equations 7,8 and 9,10 in \citet{sansom2015novel}. For $\geq$1\,g meteorite-dropping events, the following equations, modified from \citet{sansom2019determining}, were applied:
\begin{equation}
    \ln(\beta) = \ln(17.1 - 3 \ln(\alpha \sin\gamma)), \mu = 0
\label{eq:1g_oc_mu0} 
\end{equation}

\begin{equation}
    \ln(\beta) = \ln(5.7 - \ln(\alpha \sin\gamma)), \mu = 2/3
\label{eq:1g_oc_mu23} 
\end{equation}
\(\gamma\) is the entry angle of the meteoroid, and \(\mu\) is the shape change coefficient representing the rotation of a meteoroid body ($0\leq \mu \leq 2/3$). For more details into the development and applications of the \(\alpha-\beta\) methodology, we recommend reviewing the respective works \cite{gritsevich2006extra, gritsevich2007og, moreno2015new, sansom2019determining, shober2021main}. The \(\alpha-\beta\) parameters can be calculated from any event with velocity and height data and determine if a meteorite is on the ground (\url{https://github.com/desertfireballnetwork/alpha_beta_modules}).

For quantitative estimates of the final masses, assumptions need to be made regarding the drag coefficient ($C_d$), the initial cross-sectional area ($S_0$), or the initial shape coefficient ($A_0$), as well as the bulk density of the meteoroid ($\rho_m$). The atmospheric surface density ($\rho_0$) is commonly set at 1.21 kg m$^{-3}$. These assumptions are similar to those required in other methods, but the parameters in this approach generally fall within a narrow range (meteoroid densities, shape coefficients, and drag values are well-documented). Notably, $\beta$ in this methodology eliminates the need for the often uncertain assumptions about ablation parameters and luminous efficiency, which are typically difficult to determine accurately. The true strength of this method lies not in extracting individual values for these parameters but in analyzing the relationship between $\alpha$ and $\beta$. With a large dataset, patterns or groupings within these parameter spaces can reveal new insights. By rearranging the equation for $\alpha$, it becomes evident that bodies with different entry masses, angles, and volumes can produce similar $\alpha$ values, making $\alpha$ and $\beta$ more versatile and reliable predictors of meteoroid atmospheric entry outcomes than more commonly used parameter sets.

\subsection{Meteor shower identification}
To remove meteor showers from the datasets, we employed the distance function \(D_N\), which involves four geocentric quantities directly linked to observations, as proposed by \citet{valsecchi1999meteoroid}. The approach is based on the components of geocentric velocity during encounters, which are essential for \"{O}pik's theory of close encounters. Additionally, two of the new variables used in this method are near-invariant with respect to the principal secular perturbations that affect meteoroid orbits. This methodology was chosen to overcome the limitations of traditional techniques by emphasizing the use of quantities that can be computed directly from observed data rather than relying exclusively on the derivation of orbital elements.

For this study, we adopted a threshold value of 0.1 for \(D_N\) and searched for similarity to established meteor showers (\url{https://www.ta3.sk/IAUC22DB/MDC2007/Roje/roje_lista.php?corobic_roje=1&sort_roje=0}). According to the analysis conducted by \citet{shober2024generalizable} on EFN data, this threshold was demonstrated to produce less than 5\% false positives. In other words, this selected limit effectively identifies most meteoroid streams while minimizing the removal of sporadic fireballs, which can often occur due to false-positive detections — a common issue with orbital discriminants. Thus, by employing this \(D_N\) limit, we ensured that the identification of meteor showers remained robust and minimized unnecessary exclusions from our dataset. 

\subsection{Sporadic cometary component identification}
In this study, our primary goal is to characterize the atmospheric filtering effect on the impact population. Since cometary components -- whether originating from Jupiter-family comets (JFCs) or long-period comets (LPCs) -- are being filtered out by the atmosphere, their removal is necessary to better understand the atmospheric filtering effect on the sporadic asteroidal component. By isolating the asteroidal component, we can more accurately access the source regions of debris at the top of the atmosphere and which sources are too weak to survive the passage.

Meteor showers are relatively straightforward to remove from the dataset; however, identifying and removing the cometary population, particularly the JFCs and LPCs within the sporadic component, is slightly more nuanced. Traditionally, the Tisserand’s parameter ($T_J$) has been employed to differentiate cometary from asteroidal orbits, specifically to filter out JFC-related material. Yet, recent studies \citep{tancredi2014criterion,shober2021main,shoberjfcs2024} have shown that $T_J$ does not adequately discriminate JFC components, particularly within fireball databases. The percentage of fireball events originating from JFCs is estimated to be between 1\% and 5\%, reflecting the lower probability of detecting JFC debris in the sporadic population at these sizes \citep{shoberjfcs2024}. However, the proportion of fireball datasets with $2<T_J<3$ could be 25-40\%, i.e., many asteroidal sporadic are getting misclassified as JFC in origin if $T_J$ is used. 

To address this, we utilized numerical simulation results from previous in-depth studies \citep{shober2021main,shoberjfcs2024}. This previous work has already identified meteoroids detected by EFN, FRIPON, GFO, and other networks likely originating from orbits with chaotic dynamics typical of JFCs over 10,000\,yr timescales. Otherwise, for the impacts detected by EDMOND, GMN, and CAMS, which were not analysed in these previous studies, we applied the \citet{tancredi2014criterion} criterion. This more complex criterion has proven to be a significantly better method than $T_J$ for identifying objects originating from the JFC population \citep{shoberjfcs2024}. 

For LPC component identification, we retained the use of $T_J$ as a filtering criterion. Specifically, we filtered out all impacts with $T_J < 2$, as this threshold effectively excludes most asteroidal material as false classifications. Very few asteroids travel on orbits with $T_J < 2$, making this a reliable method for removing LPC-related debris from the dataset. However, for JFCs with $2 < T_J < 3$, $T_J$ is not a reliable metric for fireball data, and the \citet{tancredi2014criterion} criterion is more suitable.

By applying these refined methods, we successfully isolated the sporadic asteroidal component of our dataset. This step was crucial for understanding the physical characteristics of the meteoroids that remain after cometary material is filtered out by the atmosphere.

\subsection{Removing the Observational Bias}
Visual meteor observations are inherently constrained by the limiting magnitude of the sensors used, which varies depending on the sensor. This variation introduces a fundamental observational bias when studying the velocity or orbital distribution of fireballs and meteors. Specifically, fireball and meteor observations are biased toward detecting faster meteoroids, particularly at the smaller end of the size spectrum. This bias arises because small, slow meteoroids are not bright enough to be detected, whereas meteoroids of the same size but with larger impacting velocities remain observable. Consequently, when plotting the size frequency distribution (SFD) of observed meteoroids, a deviation from a linear relationship in log-log space becomes evident as the objects decrease in size (see Supplementary~Section~1). This deviation indicates the onset of significant observational bias in the data.

The bias leads to an overrepresentation of faster velocities and, as a result, different orbital characteristics than reality. Since this study focuses on comparing orbital distributions, it is crucial to address and mitigate this bias. While one option to correct this bias would involve weighting the observations at the smaller end of the SFD, adjusting for the overrepresentation of fast objects, such an approach introduces additional assumptions about the population distribution. We chose a more straightforward approach to avoid these assumptions and maintain simplicity in the analysis: cutting off data below the size threshold where significant observational bias begins to affect the SFD slope.

While removing a significant portion of the data in some cases, this approach avoids introducing further assumptions about the underlying population. We believe this method provides a more reliable comparison of orbital distributions, as it limits the influence of observational bias without introducing personal biases. The point at which observational bias starts to significantly affect the SFD slope can be identified in each dataset, and the data beyond that point are excluded from the analysis (see Supplementary~Figure~1). The cutoff values were determined by examining the SFD plots for each observation network.

For the CAMS, EDMOND, GMN, and EFN networks, a minimum mass cutoff of 10\,g was applied. Although all four networks can reliably detect smaller objects, we found that this tended to start to include a greater proportion of cometary material which could be accidentally included, which is not the focus of this study. FRIPON, on the other hand, was limited to a minimum starting mass of 1\,kg. Due to the hardware specifics of the FRIPON network, we found considerably more velocity bias, and a 1\,kg limit was chosen as the R-squared value for the correlations between the orbital elements and the mass became insignificant above this limit. This conservative threshold ensures that the dataset remains free from significant orbital variations due to observation bias. 

For the GFO dataset, we opted not to include it in the top-of-the-atmosphere population estimates. The GFO, which uses the same system as DFN observatories, employs a liquid crystal shutter to encode timing information into the images for precise absolute timing \citep{howie2017build,howie2017submillisecond}. However, this technique limits the detection of fireballs to those with durations longer than one second, introducing an additional bias. While it is possible to correct for this bias, it is more complex, and for the sake of maintaining precision in our analysis, we excluded the GFO data from the top-of-the-atmosphere population.

To ensure that the size range we analysed was valid and that the comparison between the top-of-the-atmosphere orbital distribution and the meteorite-dropping population was genuinely due to atmospheric selection, we also needed to verify that no significant correlations existed between mass and orbital distribution within the chosen mass range. Specifically, we wanted to ensure that the orbital characteristics of the meteoroids were not heavily influenced by their mass. We conducted this analysis for the CAMS, EDMOND, GMN, and EFN datasets. After removing contributions from JFCs, LPCs, and meteor showers, we found minimal correlation between mass and orbital distribution. Although we observed a slight correlation between smaller objects and higher eccentricity orbits, the degree of correlation was consistently below 1\%. 


\subsection{Orbital Distribution Variation and Statistical Significance}
We binned the orbital data into two-dimensional histograms for pairs of orbital elements (e.g., $a$ vs.\ $e$, $a$ vs.\ $\iota$, $\iota$ vs.\ $q$, $\iota$ vs.\ $Q$) to compare the distributions quantitatively. The bin ranges and sizes were defined to adequately cover the relevant parameter space, typically dividing each parameter range into 10 bins. For each bin, we counted the number of meteoroids from the reference population (top-of-the-atmosphere) and the subset population (meteorite-dropping events).

We performed a chi-squared test of independence to assess whether observed differences between the two populations were statistically significant in each bin. For each bin, we constructed a $2 \times 2$ contingency table with the counts of meteoroids from the reference and subset populations that fell within and outside the bin. The chi-squared statistic for each bin was calculated as:

\begin{equation}
\chi^2 = \sum_{i=1}^{2} \sum_{j=1}^{2} \frac{(O_{ij} - E_{ij})^2}{E_{ij}},
\end{equation}

where $O_{ij}$ are the observed frequencies and $E_{ij}$ are the expected frequencies under the null hypothesis of independence, calculated based on the marginal totals of the contingency table:

\begin{equation}
E_{ij} = \frac{(O_{i \cdot})(O_{\cdot j})}{N},
\end{equation}

with $O_{i \cdot}$ and $O_{\cdot j}$ representing the row and column totals, respectively, and $N$ being the total number of observations.

P-values were obtained by comparing the calculated $\chi^2$ values to the chi-squared distribution with one degree of freedom (since the contingency table is $2 \times 2$). We considered two significance levels corresponding to $2\sigma$ and $3\sigma$ confidence levels, with $\alpha = 0.0455$ and $\alpha = 0.0027$, respectively. Bins with p-values less than $\alpha$ were considered statistically significant at the corresponding confidence level.

To quantify the differences between the two populations, we calculated the relative difference in normalized densities for each bin:

\begin{equation}
\Delta = \left( \frac{f_{\text{sub}} - f_{\text{ref}}}{f_{\text{ref}}} \right) \times 100\%,
\end{equation}

where $f_{\text{sub}}$ and $f_{\text{ref}}$ are the normalized frequencies (probabilities) in the bin for the subset and reference populations, respectively. These normalized frequencies were obtained by dividing the counts in each bin by the total number of meteoroids in the respective population.

We generated density heatmaps to visualize the relative density differences across the orbital parameter space. The colour scale represented the value of $\Delta$, using a symmetric logarithmic normalization to emphasize both positive and negative differences. Bins with statistically significant differences were highlighted, with bold borders indicating significance at the $3\sigma$ level and standard borders for the $2\sigma$ level.

To interpret the observed distributions in the context of known dynamical processes, we overlaid theoretical resonance lines, such as those associated with the Kozai-Lidov mechanism. This allowed us to identify regions in orbital parameter space where dynamical resonances might influence the meteoroid population.

\subsection{NEO Clustering and Clustering Statistical Significance}
In this study, we applied a statistical significance test to investigate the degree of clustering of NEOs and fireball data. Using methods similar to those outlined in \citet{shober2024decoherence} and \citet{shober2024generalizable}, we employed a Density-Based Spatial Clustering of Applications with Noise (DBSCAN) algorithm to identify clusters in the NEO population. DBSCAN is well-suited for identifying groups of objects based on their proximity in a multi-dimensional space without specification of the number of clusters or size. For this analysis, we focused on calculating the orbital similarity between NEOs and meteoroids using the $D_H$ orbital discriminant, which considers the perihelion distance ($q$), eccentricity ($e$), inclination ($\iota$), argument of perihelion ($\omega$), and longitude of ascending node ($\Omega$) \citep{jopek1993remarks}. This method was applied to the data of 34,842 NEAs (obtained from NASA's JPL HORIZONS database), the 7,982 top-of-the-atmosphere impacts, and the 540 possible meteorite falls. 

The statistical testing was carried out using a Kernel-density estimation-based methodology developed by \citet{shober2024generalizable} to quantify the degree of similarity expected within any given random sporadic sample. By examining the cumulative D-value distributions for the NEO dataset, we observed a clear change in slope in the cumulative D-value distribution (Fig.~13 in \citealp{shober2024decoherence}), particularly around \(D_H \approx 0.01\) for only NEAs, suggesting an excess of clustering that exceeds what would be expected from random associations. After calculating the $D_H$ orbital similarity for the nearly 1 billion unique orbital pairs, the DBSCAN was applied with an epsilon ($\epsilon$) value of 0.01 and a minimum of two connections required to define a core point. This epsilon value corresponds to the observed ``kink'' location in the cumulative D-value distribution found by \citet{shober2024decoherence} for NEAs where an excess of clustering is detected beyond what would be expected from random associations within the NEA population. Please note that in \citet{shober2024decoherence}, the quoted $D_{H}$ ``kink'' value was 0.03; however, this included cometary fragments, and for NEAs, this value is less. 

By comparing the top-of-atmosphere impacts and meteorite-dropping events to these NEA clusters, we aimed to quantify the influence of tidal disruption clusters on both populations. If meteoroids from tidally-generated clusters are inherently weaker, they may be underrepresented in the meteorite falls. Indeed, while approximately 0.4\% of the top-of-the-atmosphere population belongs to one of the 11 NEA clusters identified, only about 0.2\% of the 540 potential meteorite falls meet the cluster membership criteria. This disparity suggests that meteoroids from tidally disrupted clusters, despite being common in near-Earth space, are less likely to survive the journey through the atmosphere and be recovered as meteorites.

\backmatter




\bmhead{Data availability} 
The fireball data for the top-of-the-atmosphere and falls can be accessed at \url{https://doi.org/10.5281/zenodo.14017585}. 


\bmhead{Code availability}
Available upon request from the authors. 

\bmhead{Acknowledgements}
This project has received funding from the European Union’s Horizon 2020 research and innovation programme under the Marie Skłodowska-Curie grant agreement No945298 ParisRegionFP (P.M.S.), and the grant agreement No. 101150536 (S.A.)

The Global Fireball Observatory and data pipeline is enabled by the support of the Australian Research Council (DP230100301, LE170100106). (P.M.S., H.A.R.D., S.E.D., E.K.S. P.B.)

FRIPON was initiated by funding from ANR (grant N.13-BS05-0009-03), carried by the Paris Observatory, Muséum National d’Histoire Naturelle, Paris-Saclay University and Institut Pythéas (LAM-CEREGE). VigieCiel was part of the 65 Millions d’Observateurs project, carried by the Muséum National d’Histoire Naturelle and funded by the French Investissements d’Avenir program. FRIPON data are hosted and processed at Institut Pythéas SIP (Service Informatique Pythéas), and a mirror is hosted at LTE (Le Laboratoire Temps Espace / Paris Observatory). (P.M.S., J.V., S.A., F.C., B.Z., P.V.)

\bmhead{Author contribution}
P.M. Shober collected the open-access data sources, removed the cometary components, applied the $\alpha$-$\beta$ methodology, debiased the FRIPON dataset, estimated the masses for the CAMS/EDMOND datasets, and did the statistical significance analysis of the differences between the orbital distributions. S. Anghel calculated the photometric masses for the FRIPON dataset. P.M. Shober, H. Devillepoix, J. Vaubaillon, S. Deam, and S. Anghel assisted with the data interpretation. E. Sansom, H. Devillepoix, and P. Bland facilitated the management, collection, and reduction of the GFO dataset. P.M. Shober, S. Anghel, F. Colas, J. Vaubaillon, B. Zanda, and P. Vernazza facilitated the management, collection, and reduction of the FRIPON observations. P.M. Shober, H. Devillepoix, J. Vaubaillon, S. Deam, S. Anghel, and P. Vernazza assisted with the manuscript revisions. 

\bmhead{Conflict of interest/Competing interests}
The authors declare no competing interests.



\noindent

\section*{Figure Legends}

\noindent
\textbf{Figure 1.} \textbf{Orbital distribution of 7,982 impacts detected by EDMOND, CAMS, GMN, FRIPON, and EFN networks.} Each impact is estimated to be equal to or greater than 10\,g (diameter $\gtrapprox$2\,cm) at the top of the atmosphere. Orbital distribution is normalized to the impact probability, as calculated using the method of \citet{pokorny2013opik}. Despite this, there still exists a concentration of meteoroids on orbits with q$\sim$1\,au. \\ \\ 
\textbf{Figure 2.} \textbf{Apparent initial velocity distribution for the sporadic asteroidal top-of-the-atmosphere population versus the meteorite fall populations.} The top-of-the-atmosphere population is derived from a debiased subset of observations from the EDMOND, CAMS, GMN, FRIPON, and EFN impact databases. The meteorite fall observations are taken from the GFO, EFN, and FRIPON datasets. \\ \\
\textbf{Figure 3.} \textbf{Orbital density heatmaps comparing sporadic $>10$\,g top-of-atmosphere asteroidal impacts and $>1$\,g meteorite falls.} Panels show: (a) semi-major axis vs eccentricity, (b) semi-major axis vs inclination, (c) inclination vs perihelion distance, and (d) inclination vs aphelion distance. The colours indicate percentage differences in normalized density (red = overrepresented in the meteorite population; blue = underrepresented in the meteorite population). Statistical significance (chi-squared test) is highlighted by black bins at the 3$\sigma$ level and all others at 2$\sigma$. The dashed lines mark the perihelion evolution ($q$) as a function of inclination ($i$) under Kozai-Lidov resonance for orbits with $a=2.5$\,au. \\
\ldots



\end{document}


\title[Supplementary Information]{Supplementary Information}

\maketitle

\tableofcontents





\section{Top-of-atmosphere Data Selection and Debiasing}\label{secA1}

\begin{figure} 
\centering \includegraphics[width=1\linewidth]{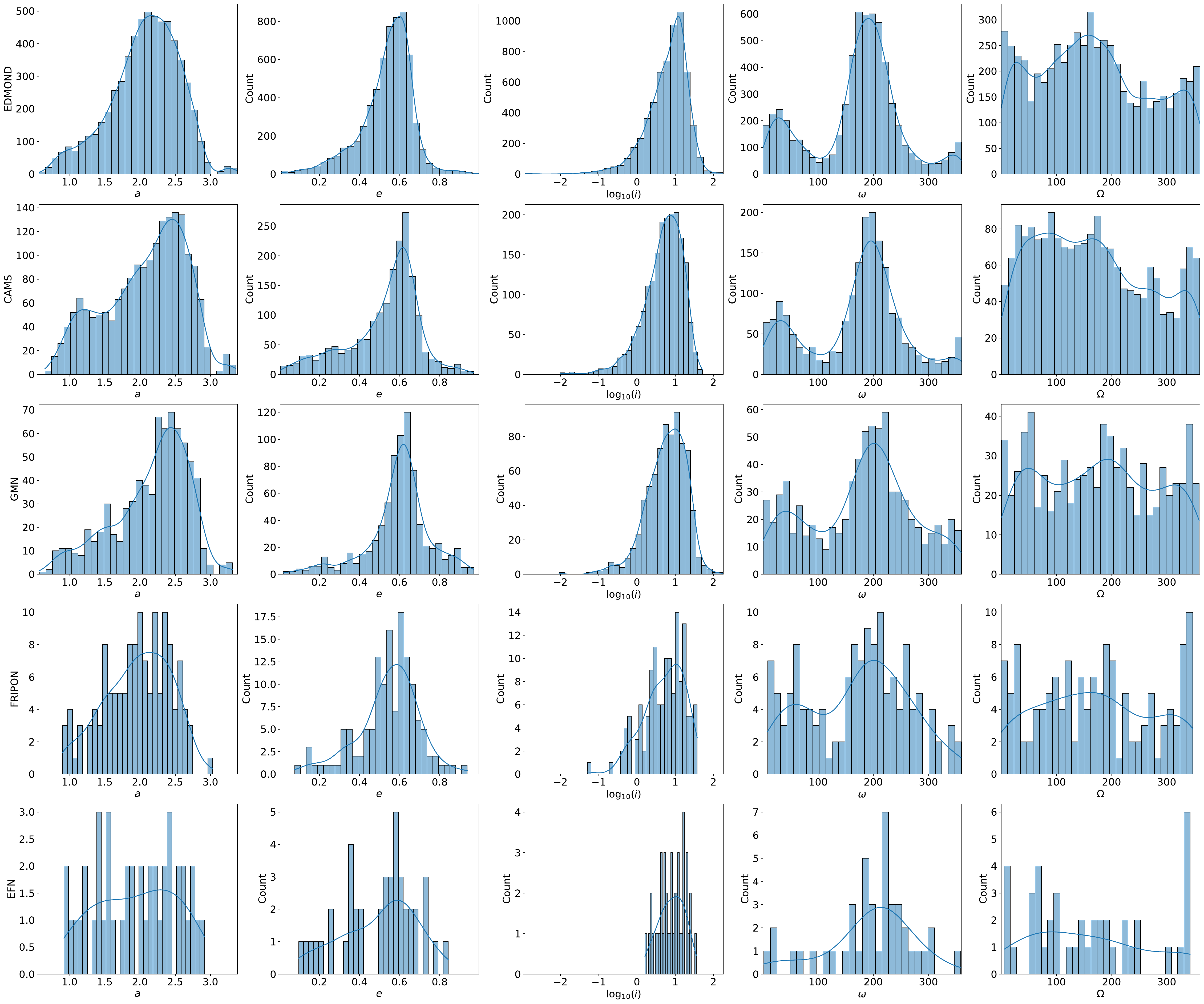} 
\caption{Orbital distributions for the 7,982 meteoroid impacts detected by the EDMOND, CAMS, GMN, FRIPON, and EFN networks. Each row represents the orbital distribution from a single network, showing general similarities and consistent trends across the datasets. Minor variations may exist due to network-specific observational biases or data processing methods.} 
\label{fig:sup_all_distrobutions_top} 
\end{figure}

Supplementary~Figure~\ref{fig:sup_all_distrobutions_top} shows the orbital distributions for all the data used by the EDMOND, CAMS, GMN, FRIPON, and EFN networks for the top-of-the-atmosphere. This dataset comprises the 7,982 impacts detected by these networks, each estimated to have a minimum initial mass greater than 10\,g (except for FRIPON, as discussed below). The distributions exhibit general similarities across all networks, indicating consistent trends in the top-of-the-atmosphere meteoroid population. Some variations might exist due to minor observational biases not being accounted for, specifics to each network, or the processing of publicly available data (e.g., EDMOND, CAMS, GMN). The size-frequency distribution (SFD) for the FRIPON dataset shows deviation from linearity at sub-kilogram masses due to observational biases at smaller masses. We found an R-squared value of 0.29 when performing a linear regression between mass and eccentricity for FRIPON meteoroids $>$\,10\,g, with a P-value approaching zero. This indicates a significant correlation between mass and orbital parameters in the FRIPON dataset. In other words, smaller masses are more likely to have higher eccentricities, suggesting that observational biases influence the dataset more than others in this mass range. Higher-velocity meteoroids (which tend to have higher eccentricities) are easier to detect for smaller meteoroids, leading to an overrepresentation of small, high-eccentricity meteoroids in the FRIPON data. In contrast, the R-squared value for the same linear regression between mass and eccentricity for the EFN dataset was found to be only 0.05 in the same mass range.

Although there are slight differences between the mass estimates from different networks, most of these differences seem to be due to this velocity observation bias. All the other networks are not significantly biased above 10 grams. However, the mass estimates for large GMN and EDMOND meteoroids are extremely overestimated due to camera sensor saturation. In this study, we do not examine these masses distribution themselves, so their overestimation does not affect our results, as we are primarily concerned with the correlation between mass and orbital parameters.


\section{Mass-dependent orbital variations}\label{secA2}%
This appendix examines our datasets' potential correlations between meteoroid mass and orbital elements. This analysis is crucial to ensure that our comparison of orbital distributions is not biased by mass-dependent effects.

\begin{figure}
    \centering
    \includegraphics[width=1\linewidth]{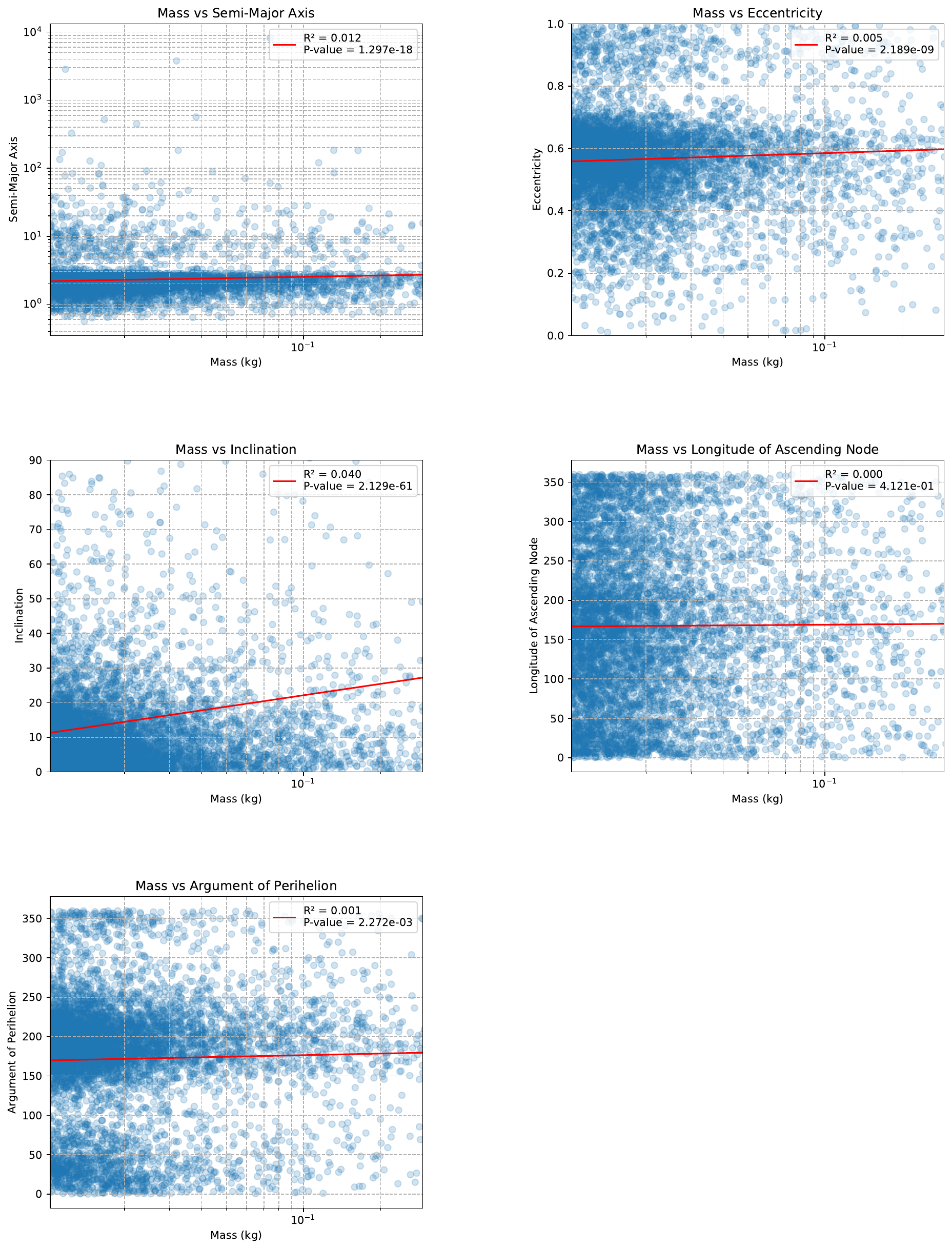}
    \caption{Correlation between meteoroid mass and orbital elements for the $>$\,10\,g EDMOND dataset. Linear regression lines are shown with $R^2$ and p-values. Long-period and Jupiter-family comet contributions are removed (See Methods).}
    \label{fig:supp_EDMOND_correlation}
\end{figure}

\begin{figure}
    \centering
    \includegraphics[width=1\linewidth]{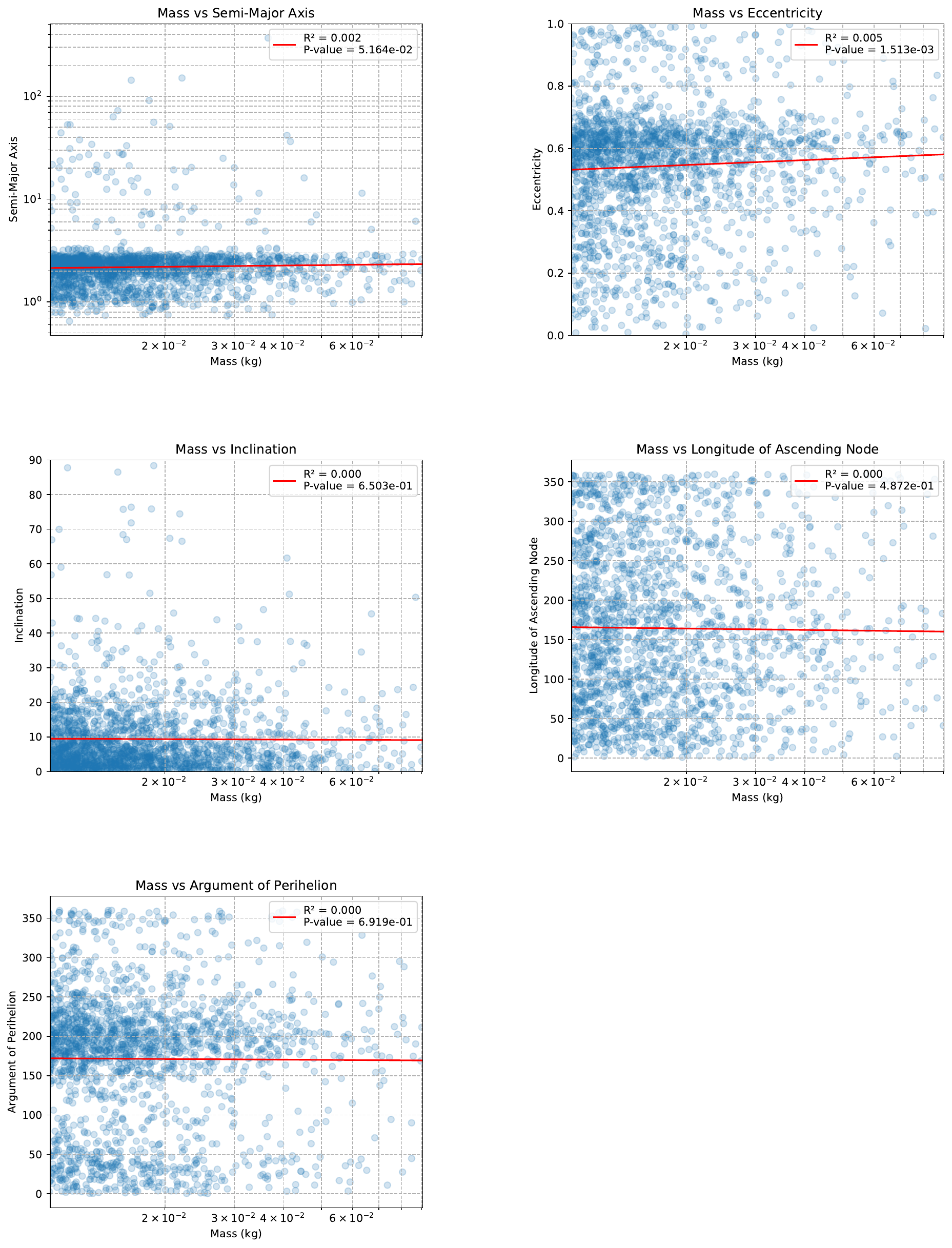}
    \caption{Correlation between meteoroid mass and orbital elements for the $>$\,10\,g CAMS dataset. Linear regression lines are shown with $R^2$ and p-values. Long-period and Jupiter-family comet contributions are removed (See Methods).}
    \label{fig:supp_CAMS_correlation}
\end{figure}

\begin{figure}
    \centering
    \includegraphics[width=1\linewidth]{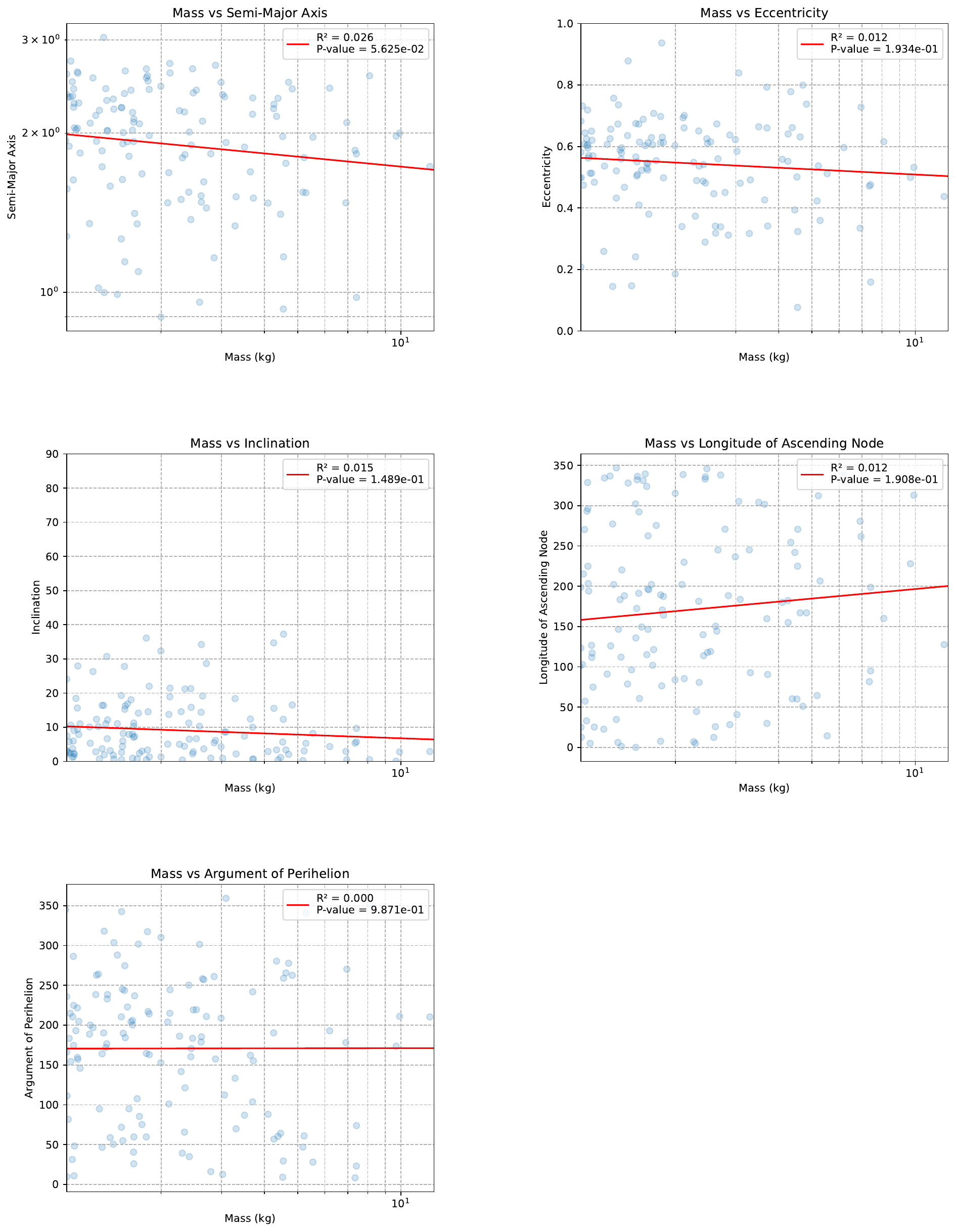}
    \caption{Correlation between meteoroid mass and orbital elements for the $>$\,1\,kg FRIPON dataset. Linear regression lines are shown with $R^2$ and p-values. Long-period and Jupiter-family comet contributions are removed (See Methods).}
    \label{fig:supp_FRIPON_correlation}
\end{figure}

\begin{figure}
    \centering
    \includegraphics[width=1\linewidth]{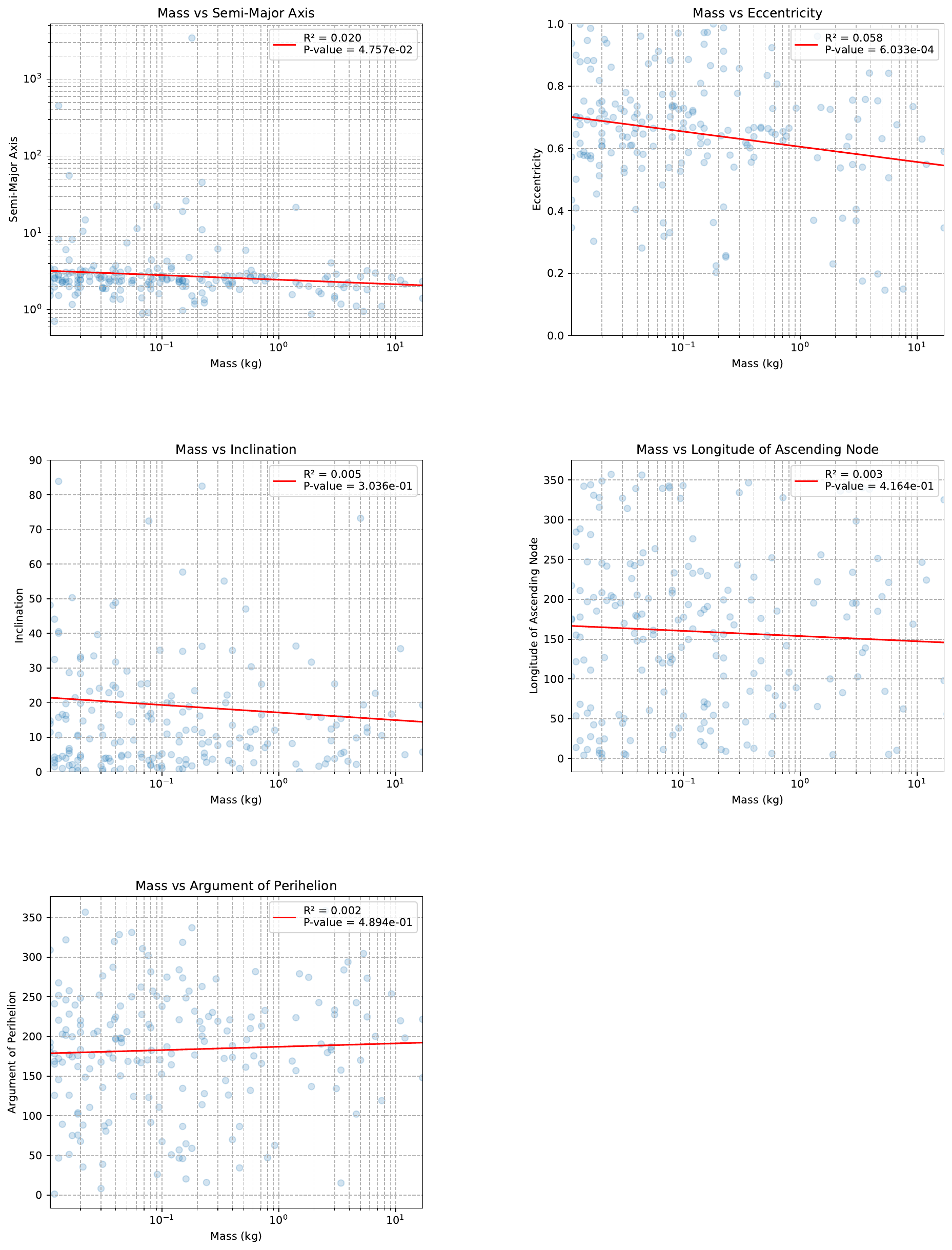}
    \caption{Correlation between meteoroid mass and orbital elements for the $>$\,10\,g EFN dataset. Linear regression lines are shown with $R^2$ and p-values. Long-period and Jupiter-family comet contributions are removed (See Methods).}
    \label{fig:supp_EFN_correlation}
\end{figure}

\begin{figure}
    \centering
    \includegraphics[width=1\linewidth]{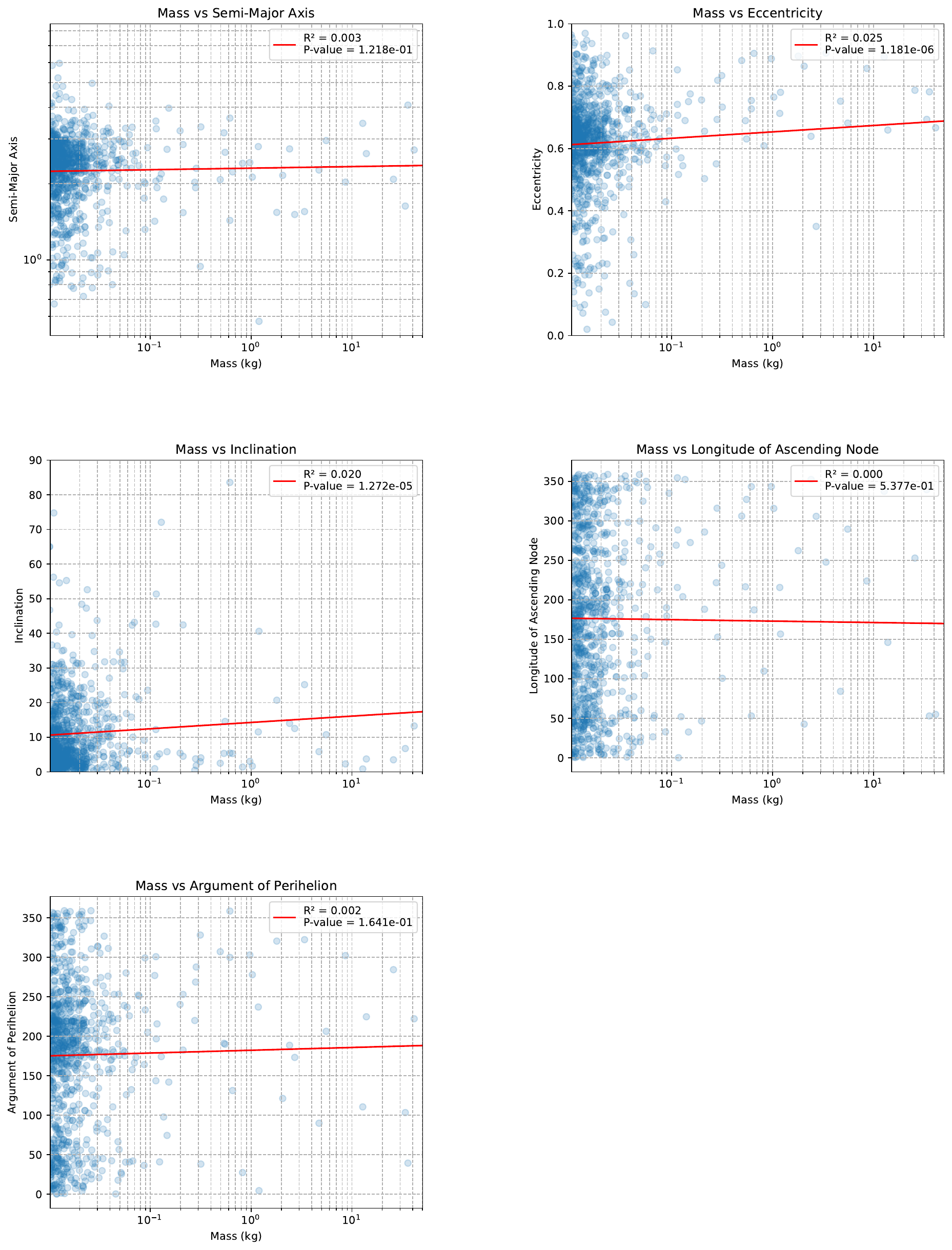}
    \caption{Correlation between meteoroid mass and orbital elements for the $>$\,10\,g GMN dataset. Linear regression lines are shown with $R^2$ and p-values. Long-period and Jupiter-family comet contributions are removed (See Methods).}
    \label{fig:supp_GMN_correlation}
\end{figure}

Figs.\ref{fig:supp_EDMOND_correlation}-\ref{fig:supp_GMN_correlation} show the relationships between meteoroid mass and various orbital elements for each network dataset used for the top-of-the-atmosphere population. A linear regression was performed for each plot, and the coefficient of determination ($R^2$) and p-value were calculated to assess the strength and significance of any correlations.

The coefficient of determination, $R^2$, quantifies the proportion of variance in the dependent variable (mass) that is predictable from the independent variable (orbital element). An $R^2$ value of 1 indicates a perfect correlation, while an $R^2$ value of 0 indicates no correlation. In nearly all cases, the $R^2$ values are well below 0.05 and, in most cases, nearly zero. These low $R^2$ indicate that there is no significant correlation between meteoroid mass and orbital elements in the datasets used for this study.

However, it is important to note that the datasets used in this study to estimate the top-of-the-atmosphere population do not contain many metre-scale impactors. According to publicly available data from U.S. Government sensors on the CNEOS website\footnote{\url{https://cneos.jpl.nasa.gov/fireballs/}}, the orbital distribution of metre-scale impacts appears to be significantly different from that of smaller meteoroids. This observation is intriguing, as it suggests there may be orbital variability at these larger sizes, possibly due to different source regions or dynamical/physical processes affecting larger objects. Nevertheless, considering that numerous studies have highlighted the dubiousness of the CNEOS dataset's orbits and it's lack of formal uncertainties \citep{devillepoix2019observation}, these observations need to be confirmed with more open and reliable data sources.

\section{Meteorite fall identification using $\alpha$-$\beta$}\label{secA3}%
To identify meteorite-dropping fireballs, we employed the $\alpha$-$\beta$ criterion introduced by \citet{sansom2019determining}, which offers a more elegant and physically meaningful approach compared to traditional final height and velocity limits. The $\alpha$ and $\beta$ parameters are dimensionless coefficients that represent the ballistic coefficient and mass-loss parameter, respectively, encapsulating the physical properties of the meteoroid. This method can be applied to any fireball network and has proven effective for holistic meteorite fall identification using only velocity and height data. We also applied a limit of at least 20\% deceleration observed during the meteoroid's atmospheric entry.

We applied the $\alpha$-$\beta$ methodology to data from FRIPON and GFO networks to estimate the meteorite fall population's orbital distribution. For EFN events, we used the final mass estimates provided by \citet{borovivcka2022_one}. In our analysis, we considered different limiting masses (e.g., 50 grams and 1 kilogram) to confirm the observed trends. The limiting lines on the $\alpha$-$\beta$ plots correspond to the threshold for meteorite survival, shifting downward for larger masses, thus reducing the number of objects falling below the curve. The bold red line in the Supplementary~Figure~\ref{fig:supp_alpha_beta_GFO},\ref{fig:supp_alpha_beta_FRIPON} represents the carbonaceous chondrite curve for likely meteorite droppers, while the normal red line indicates possible carbonaceous droppers. Similarly, the bold black line denotes the likely ordinary chondrite droppers, and the regular black line represents possible ordinary chondrite droppers. This classification aids in distinguishing between different meteoroid compositions based on their atmospheric behaviour. The equations for the lines were derived by modifying Eq.\,6 from \citet{sansom2019determining}, supplementing in the appropriate final mass and density values.  

\begin{figure}
    \centering
    \includegraphics[width=1\linewidth]{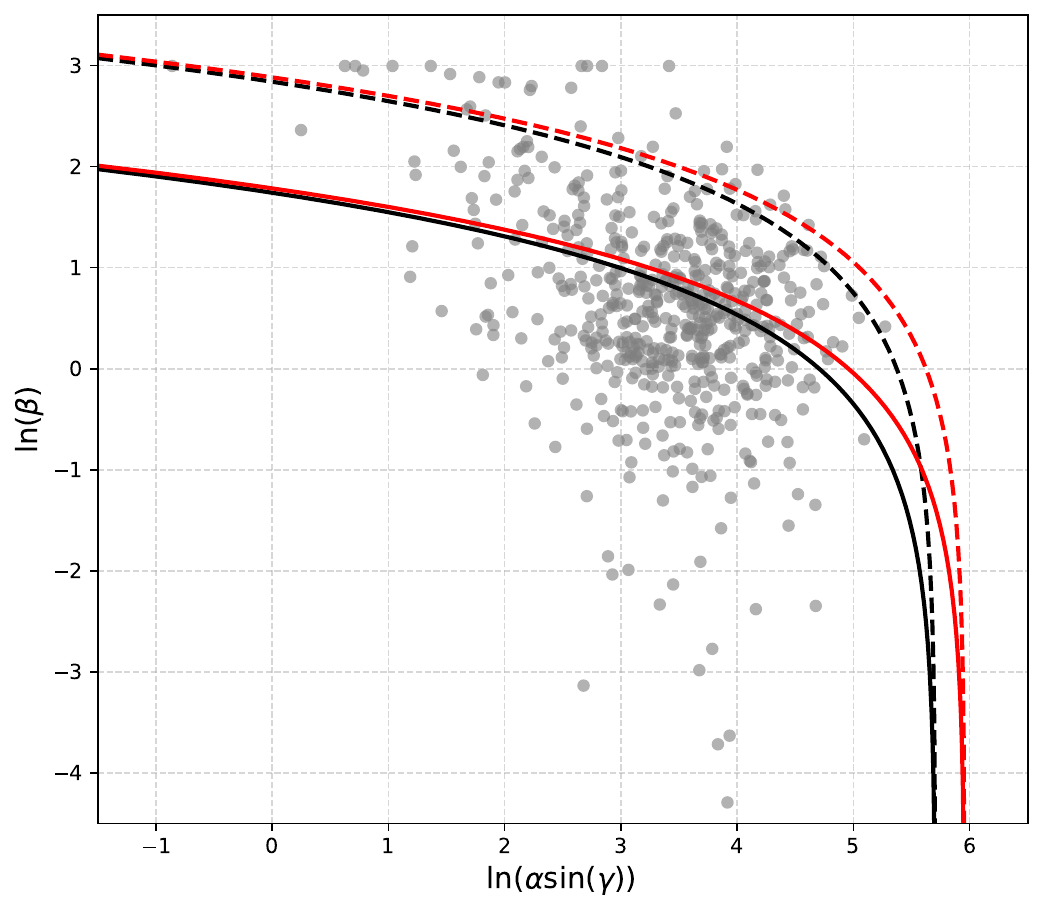}
    \caption{Distribution of fireballs observed by the GFO network in the $\alpha$-$\beta$ parameter space, with trajectory slope dependence removed. The bounding lines represent the limiting values for which a 1\,g meteorite could possibly be dropped; events below the lines indicate potential meteorite-dropping fireballs. Bold lines correspond to a shape change coefficient $\mu = 2/3$ (uniform ablation across the surface, a meteoroid is spinning), and thin lines correspond to $\mu = 0$ (no spin). Black lines represent ordinary chondrite densities ($\rho = 3500$,kg,m$^{-3}$), and red lines represent carbonaceous chondrite densities ($\rho = 2400$,kg,m$^{-3}$). Bold lines indicate thresholds for likely meteorite droppers, while regular lines indicate possible droppers.}
    \label{fig:supp_alpha_beta_GFO}
\end{figure}

\begin{figure}
    \centering
    \includegraphics[width=1\linewidth]{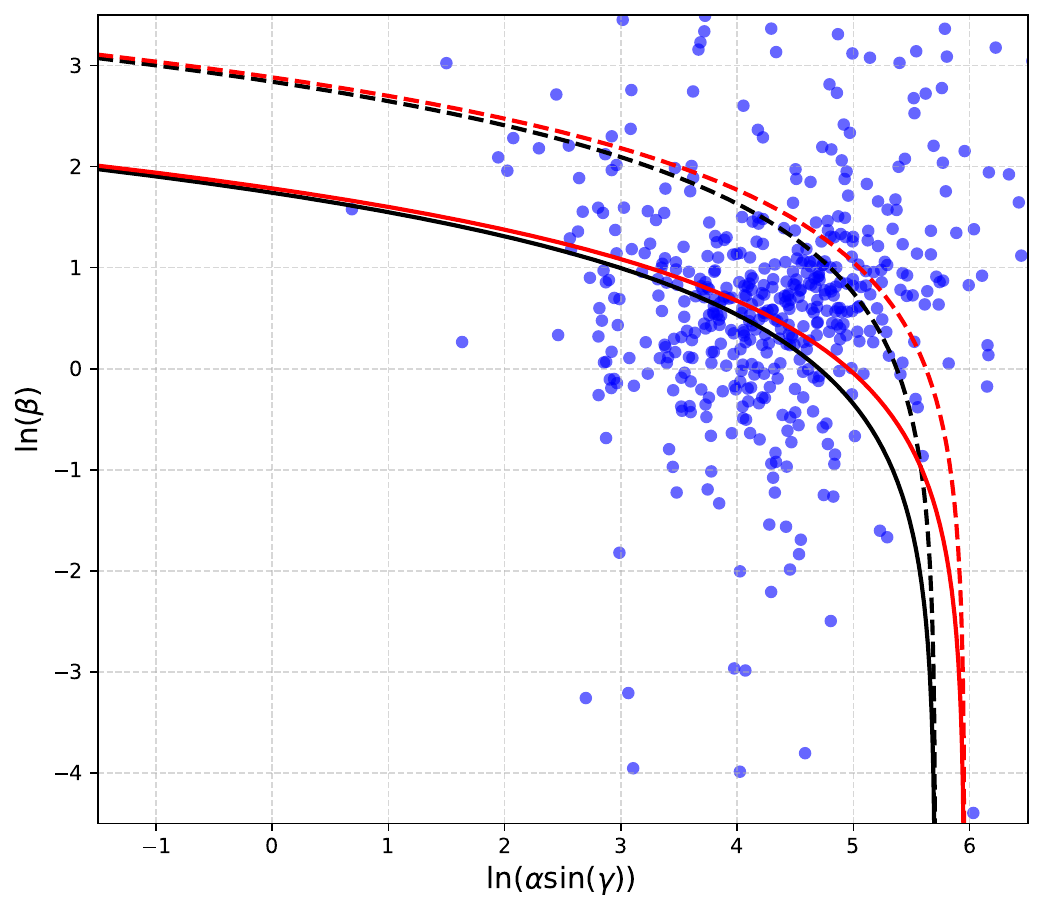}
    \caption{Distribution of fireballs observed by the FRIPON network in the $\alpha$-$\beta$ parameter space, with trajectory slope dependence removed. The bounding lines represent the limiting values for which a 1\,g meteorite could be dropped; events below the lines indicate potential meteorite-dropping fireballs. Bold lines correspond to a shape change coefficient $\mu = 2/3$ (uniform ablation across the surface, a meteoroid is spinning), and thin lines correspond to $\mu = 0$ (no spin). Black lines represent ordinary chondrite densities ($\rho = 3500$,kg,m$^{-3}$), and red lines represent carbonaceous chondrite densities ($\rho = 2400$,kg,m$^{-3}$). Bold lines indicate thresholds for likely meteorite droppers, while regular lines indicate possible droppers.}
    \label{fig:supp_alpha_beta_FRIPON}
\end{figure}

\section{Meteorite Survival Rate}\label{secA4}%
\begin{figure}
    \centering
    \includegraphics[width=1\linewidth]{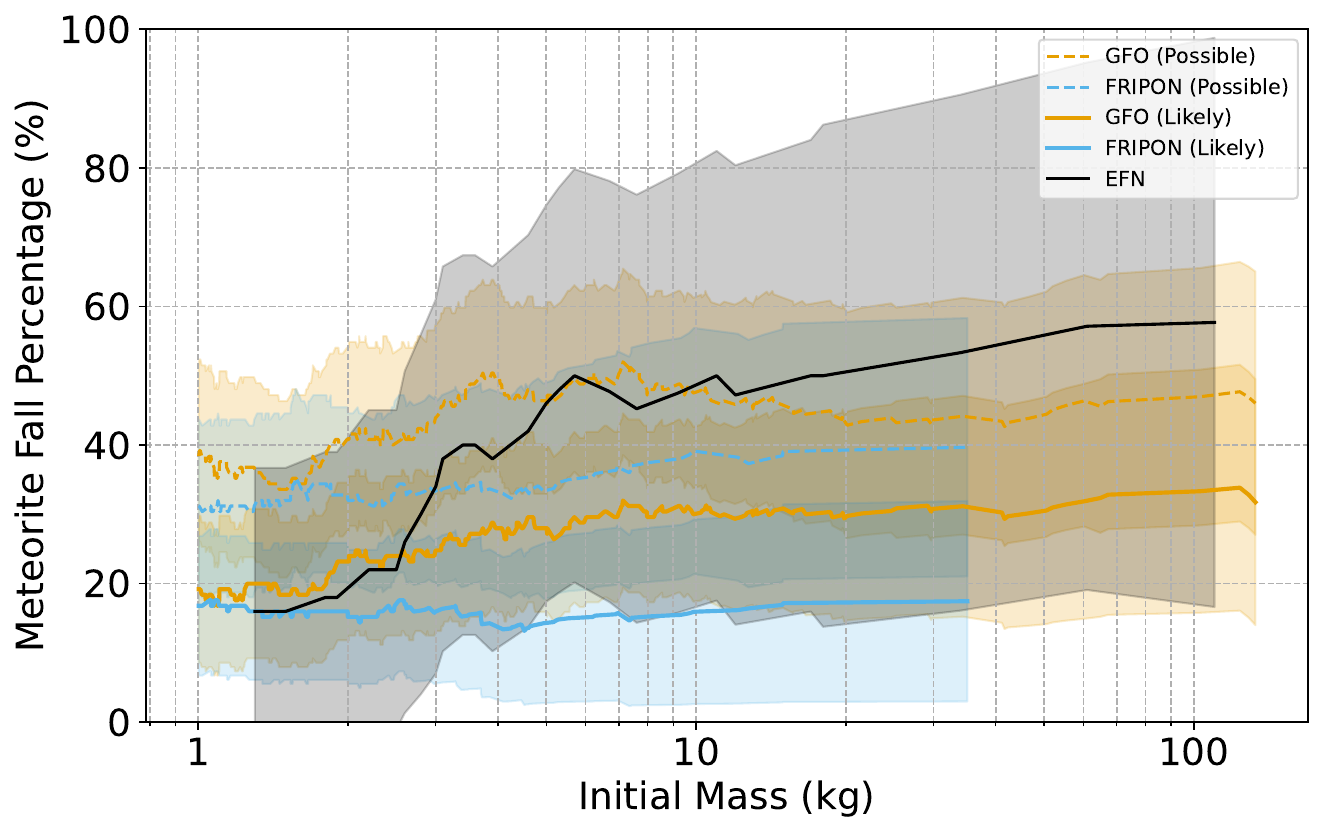}
    \caption{Estimated percentage of meteorite-dropping events as a function of meteoroid mass for the GFO, FRIPON, and EFN networks. The shaded areas represent the 3$\sigma$ uncertainties where the line indicates the nominal estimate. GFO and FRIPON events are separated into likely and possible droppers based on the $\alpha$-$\beta$ criterion, considering meteorites with a final mass of at least 1\,g.}
    \label{fig:survival_percentage_by_mass}
\end{figure}

Despite the limited dataset of 540 events, which introduces uncertainties, we observe that for objects greater than 1 kilogram, the survival rates tend to concentrate between 35\% and 55\% for mean values in the mass range between 1\,kg and 100\,kg. However, due to the small sample size, more data are needed to confirm these trends conclusively. Supplementary~Figure~\ref{fig:survival_percentage_by_mass} illustrates the percentage of meteorite droppers recorded by the GFO, FRIPON, and EFN networks as a function of the initial estimated mass. The shaded areas correspond to the 3$\sigma$ uncertainties of the estimates. In this analysis, we have separated the GFO and FRIPON events into likely and possible droppers based on the $\alpha$-$\beta$ criterion, focusing on meteorites with a final mass of at least 1\,g.

Rolling averages and standard errors were computed over mass bins to account for the limited number of events and to smooth the data. This allowed us to estimate the survival rates and their uncertainties. 

\section{Historical Minimum perihelion based on Orbit}\label{secA5}%

To investigate the influence of perihelion history on meteoroid survivability, we utilized the lookup table from \citet{toliou2021minimum}, which provides probabilistic assessments of the minimum perihelion distances (q) and associated dwell times for NEAs based on their orbital elements and absolute magnitudes (H). This lookup table is derived from the NEA population model by \citet{granvik2018debiased} and accounts for the super-catastrophic disruption of asteroids at small q due to solar irradiation \citep{granvik2016super}.

\begin{figure}
    \centering
    \includegraphics[width=1\linewidth]{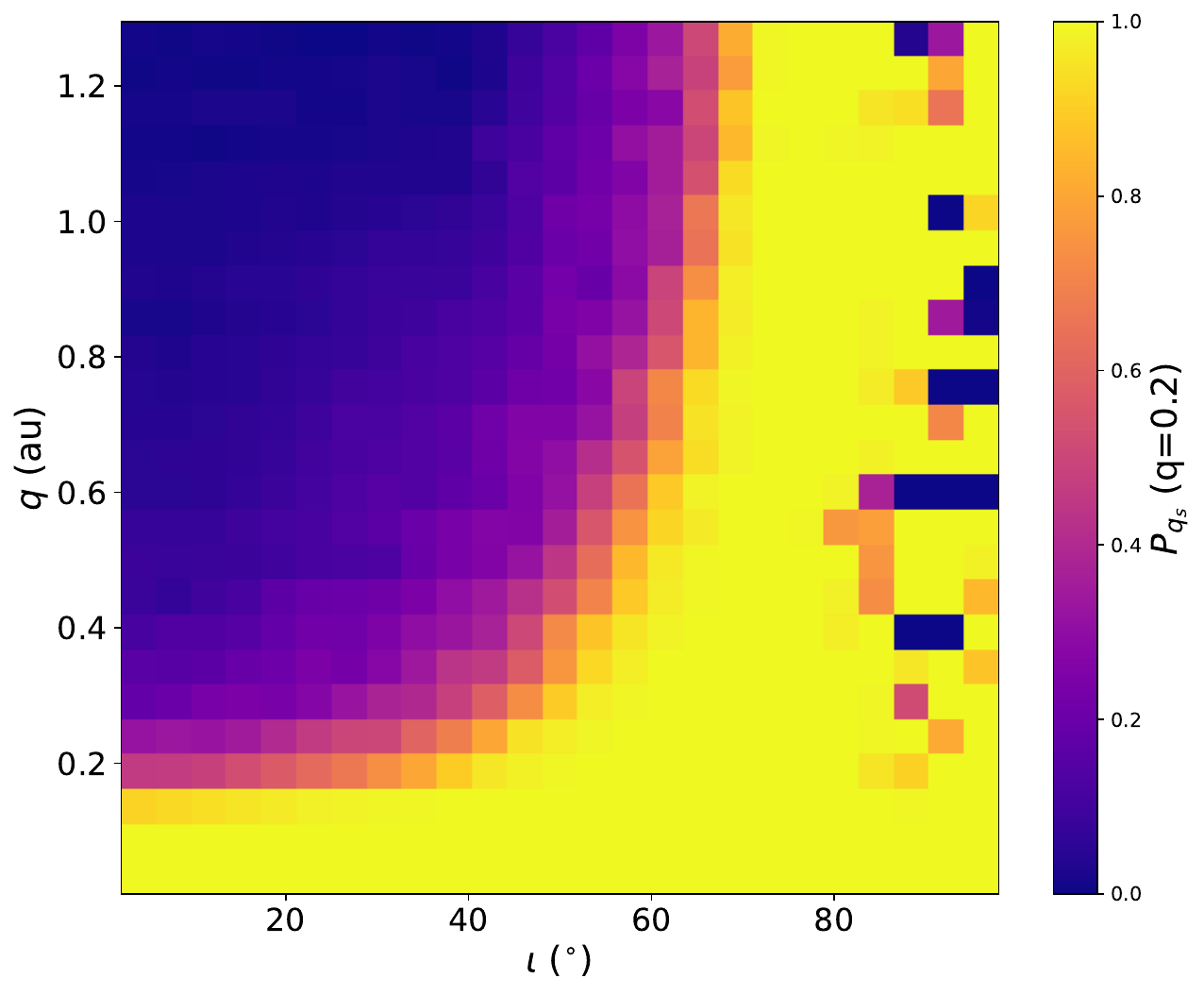}
    \caption{Heat map of the percentage of NEAs that reach $q<$\,0.2\,au during their lifetimes. The colour indicated the percentage of asteroids predicted by the Granvik et al.\ (2018) NEA population model to have reached a minimum perihelion distance of $q<$\,0.2\,au, as a function of their current inclination ($\iota$) and perihelion distance ($q$), based on the lookup table from \citet{toliou2021minimum}.}
    \label{fig:toliou_q_value_heat_map}
\end{figure}

Using this resource, we generated a heat map (Supplementary~Figure~\ref{fig:toliou_q_value_heat_map}) showing the percentage of asteroids ($15<H<25$)predicted to have reached a minimum perihelion distance of $q<$\,0.2\,au as a function of their current inclination ($\iota$) and perihelion distance. The trend observed in the heat map identically matches our survival heat maps derived from meteoroid data, indicating that perihelion history is the key factor in meteoroid survivability.

The results also confirm that meteoroids currently observed with $q\sim$\,1\,au have not previously experienced closer approaches to the Sun. Weak meteoroids with current perihelion distances near 1,au have avoided the intense solar heating.

\bibliography{sn-bibliography}